\documentclass[english,aps,prper,reprint,showpacs,titlepage,longbibliography]{revtex4-2}   

\usepackage{soul}
\usepackage{xcolor}
\soulregister{\textit}{1}

\definecolor{lightgray}{gray}{0.9}

\sethlcolor{lightgray}
\usepackage[utf8]{inputenc}
\usepackage[compatibility=false]{caption} 
\usepackage{rotating} 
\usepackage{booktabs} 
\usepackage{tabularx} 
\usepackage{ragged2e} 
\usepackage{caption} 
\usepackage[T1]{fontenc}	
\usepackage{geometry}
\usepackage{times}
\usepackage{hyperref}  
\hypersetup{colorlinks=true,urlcolor=blue,citecolor=blue,linkcolor=blue} 
\usepackage{array}
\usepackage{enumerate}
\usepackage{enumitem}
\usepackage{amsmath}
\usepackage{caption}
\usepackage{amssymb}
\usepackage{tikz}
\usepackage{graphicx}
\usepackage{multirow}
\usepackage{adjustbox} 
\usepackage{tcolorbox}
\usepackage{xcolor}  
\definecolor{lightgreen}{rgb}{0.56, 0.93, 0.56}  
\usepackage{ragged2e}
\usepackage{float}
\usepackage[utf8]{inputenc}

\usepackage[normalem]{ulem}
\usepackage{setspace} 
\usepackage{placeins} 
\usepackage{adjustbox} 
\usepackage{soul}   
\usepackage{xcolor} 
\definecolor{lightgreen}{rgb}{0.56, 0.93, 0.56}  
\sethlcolor{lightgreen}  

\begin{document}
\begin{titlepage}

\title{Presenting a STEM Ways of Thinking Framework for Engineering Design-based Physics Problems}

 \author{Ravishankar Chatta Subramaniam}
 \affiliation{Department of Physics and Astronomy, Purdue University, 525 Northwestern Ave, West Lafayette, IN 47907, U.S.A.} 

 \author{Jason W. Morphew}
 \affiliation{School of Engineering Education, Purdue University, 610 Purdue Mall, West Lafayette, IN 47907, U.S.A.}
   
 \author{Carina M. Rebello}
 \affiliation{Dept. of Physics, Toronto Metropolitan University, 350 Victoria Street, Toronto, ON M5B 2K3, Canada}
  
 \author{N. Sanjay Rebello}
 \affiliation{Dept. of Physics and Astronomy / Dept. of Curriculum \& Instruction, Purdue University, West Lafayette, IN 47907, U.S.A.} 


\begin{abstract}
Investigating students' thinking in  classroom tasks, particularly in science and engineering, is essential for improving educational practices and advancing student learning. In this context, the notion of \textit{Ways of Thinking} (WoT) has gained traction in STEM education, offering a framework to explore how students approach and solve interdisciplinary problems. Building on our earlier studies and contributing to ongoing discussions on WoT frameworks, this paper introduces a new WoT framework—Ways of Thinking in Engineering Design-based Physics (WoT4EDP). WoT4EDP integrates five key elements—design, science, mathematics, metacognitive reflection, and computational thinking—within an undergraduate introductory physics laboratory. This novel framework highlights how these interconnected elements foster deeper learning and holistic problem-solving in ED-based projects. A key takeaway is that this framework serves as a practical tool for educators and researchers to design, implement, and analyze interdisciplinary STEM activities in physics classrooms. We describe the development of WoT4EDP, situate it within undergraduate STEM education, and characterize its components in detail. Additionally, we compare WoT4EDP with two contemporary frameworks—Dalal {\em et al.} (2021) and English (2023)—to glean insights that enhance its application and promote interdisciplinary thinking. This paper is the first of a two-part series. In the upcoming second part, we will demonstrate the application of the WoT4EDP framework, showcasing how it can be used to analyze student thinking in real-world, ED-based physics projects.

    \clearpage
  \end{abstract}

\maketitle
\end{titlepage}
\maketitle
\section{Introduction}

Few would dispute the assertion that science education is continually evolving, and that undergraduate science education is in constant need of reform. The history of science education has reinforced the need to constantly adapt to and incorporate new discoveries in science and technology~\cite{sirnoorkar2024}, reflect on and address the ever-expanding educational challenges, and transition towards creating more active, student-centered learning environments~\cite{laws1996undergraduate, hilborn1997guest, pollock2008sustaining, otero2017past,laverty2022new, brookes_etkina_2020}. 

Among the several innovations in educational practices is the approach of teaching STEM (science, technology, engineering, and mathematics) by integrating multiple disciplines~\cite{kelley2016conceptual, roehrig2021beyond}. As educators and researchers, we have a responsibility towards preparing students in being~\textit{``well informed about technical matters and well educated in the STEM subjects''}~\cite[preface.vii]{national2013adapting}. Students need to be provided both opportunities and guidance to develop a variety of thinking skills. In this context, the notion of \textit{Ways of Thinking} (WoT) has gained traction in STEM education. A few researchers~\cite{dalal2021developing, slavit2019stem, slavit2021student, english2023ways} have even proposed WoT frameworks in the context of STEM-based problem solving and Engineering Education Research. Our goal in this paper is to propose our own WoT framework, designed specifically for Engineering Design (ED)-based physics learning.

Our confidence in presenting a fresh framework stems from a series of prior qualitative studies~\cite{subramaniam2023narst, ravi_perc_2023} on the \textit{design-science gap}~\cite{vattam2008foundations, kolod_punt_2003, chao2017bridging} in engineering design~(ED)-based~\cite{capobianco2018characterizing} projects in an introductory physics laboratory. Most notably, in our recent paper~\cite{ravi_prper_2024}, we had presented an elaborate analytic framework to explore, examine, and operationalize \textit{design thinking} and \textit{science thinking}, drawing heavily from various qualitative research methodologies. We investigated how students engaged in design thinking, science thinking, mathematical thinking, and metacognitive reflection while solving an instructor-assigned ED problem. On the basis of our analysis, we suggested transitioning from the deficit-framed  notion of design-science \textit{gap} to the asset-framed notion of design-science \textit{connection} (see Table~\ref{tab:data_earlier_studies}). 

Continuing on our explorations and drawing inspiration from the works of English~\cite{english2023ways}, Dalal {\em et al.}~\cite{dalal2021developing}, and Slavit {\em et al.}~\cite{slavit2019stem, slavit2021student}, we now propose a novel STEM \textit{Ways of Thinking} framework tailored for Engineering Design (ED)-based physics contexts, wherein students engage in projects that integrate physics, mathematics, and design thinking to solve real-world-inspired challenges. This framework—referred to as Ways of Thinking for Engineering Design-based Physics (WoT4EDP) throughout this paper—offers a structured approach to investigating and facilitating students' thinking and learning in undergraduate physics education. WoT4EDP addresses the complexities of integrating design and science, providing a detailed analytical lens to examine how students apply physics principles, utilize mathematical reasoning, employ computational tools, and engage in reflective thinking as they iterate on their design to develop practical, effective solutions.

In line with the definitions proposed by Dalal {\em et al.}~\cite[p.109]{dalal2021developing} and English~\cite[p.1220]{english2023ways}, we define \textit{Ways of Thinking} in the context of ED-based projects in physics as the approaches students adopt to intentionally think, act, and engage throughout their design projects, guiding their decision-making processes. These interconnected ways of thinking constitute a comprehensive framework for exploring, analyzing, evaluating, reflecting, and solving complex, interdisciplinary challenges with a focus on practical solutions. In our context, \textit{Ways of Thinking} illuminate how students approach engineering design problems, apply physics concepts, make decisions, and engage in iterative processes to develop conceptual solutions and deepen their understanding. Through this study, we aim to introduce a \textit{Ways of Thinking} framework into Physics Education Research (PER), and in the process extend the discussion on development and significance of such frameworks. 

While elaborating on the elements of our framework, we draw upon a range of reform documents~\cite{AAPT_2023_lab_reco, NGSS2013, NRC2012, nrc_dber_2012, math_cupm_2015, nrc_CT_2010} with particular emphasis on physics education and PER. Based on our experience both as educators and researchers, we are confident our framework, in addition to extending the discourse on WoT frameworks, offers valuable insights for multiple stakeholders: educators, researchers, and students. WoT frameworks depend on educational contexts which, arguably, are inherently complex. It is essential that we remain open to, and learn from, a spectrum of perspectives. To this end, we engage in a detailed comparison with two recent contemporary WoT frameworks. 
In this study, we have two specific research goals (RGs). 
\vspace{0.2cm}

\begin{minipage}{0.90\linewidth}
    \textbf{RG1:} \textit{Introduce our framework, describe its development, and characterize each of its elements within the context of undergraduate STEM education.}
    \label{RG1}
\end{minipage}

\vspace{0.2cm}
\begin{minipage}{0.90\linewidth}
    \textbf{RG2:} \textit{Compare our proposed framework with two other recently developed frameworks: one by Dalal {\em et al.}~\cite{dalal2021developing} and another by English~\cite{english2023ways}.}
    \label{RG2}
\end{minipage}
\\

This theoretical paper is the first in a two-part series. Here, we focus on the conceptual development and comparative analysis of the WoT4EDP framework, grounding it in reform documents, relevant research, and our prior studies.

In the follow-up article, extending our earlier study~\cite{ravi_perc_2024}, we will apply the framework to new student-generated data to examine how students approach and solve ED problems. By presenting a transparent and rigorous account of our qualitative analysis, we aim to advance discourse on qualitative methodologies in Physics Education Research.

\section{Literature Review}\label{sec:II}
\subsection{The National Research Council report - 2013}\label{sec:II.A}
The National Research Council (NRC), in its 2013 report~\textit{Adapting to a Changing World: Challenges and Opportunities in Undergraduate Physics Education} highlights the unique role of physics in shaping students' minds and society by noting:  

\begin{quote}
\textit{``Undergraduate physics education provides students with unique skills and ways of thinking that are of profound value to the students and to society''}~\cite[Chapter 1, p.17]{national2013adapting}.
\end{quote}

Langenberg, the committee chair, was profuse in his appreciation of the members for their \textit{``deep dedication to physics and the ways of thinking that characterize it''}~\cite[Preface, p.viii]{national2013adapting}. The report, while making recommendations to improve undergraduate physics education, goes on to urge educators~\textit{``to actively engage students in the learning process, paying attention to their spontaneous ways of thinking...''}~\cite[Chapter 3, p.75]{national2013adapting}. The lead author of the current study drew significant inspiration from the emphasis on the words \textit{characterize} and \textit{spontaneous}, and the repeated use of the phrase \textit{ways of thinking} which appears at least seven times in the report. However, the report does not operationally define \textit{ways of thinking}.

\subsection{`Ways of Thinking' - More Than Just A Phrase}\label{sec:II.B}
Even though the NRC report \cite{national2013adapting}, has not defined \textit{ways of thinking}, several researchers have explored the notion of \textit{Ways of Thinking} in STEM and Engineering Education Research (EER) contexts. Denick and colleagues~\cite{denick2013stem} identified ways of thinking such as science thinking, technology thinking, engineering thinking, and mathematical thinking. Acknowledging that identifying strict boundaries between these ways of thinking would be challenging, they recommend a holistic integration of these ways of thinking to promote student learning. Notably, in addition to discussing \textit{science thinking}, they also reference \textit{engineering design thinking}, both of which were explored in considerable depth in our recent studies~\cite{subramaniam2023narst, ravi_perc_2024, ravi_prper_2024}. 

It is our considered view that educators must feel empowered to interpret and apply the notion of \textit{Ways of Thinking} in ways that best resonate with their individual teaching styles, unique educational contexts and objectives, disciplinary practices, and the specific needs of their students. By adopting a flexible approach, educators can better support diverse learning outcomes and promote a deeper understanding of, and integration across, STEM disciplines.

\subsection{STEM Ways of Thinking (SWoT) - The Origins}\label{sec:II.C}
From our literature search, driven by a curiosity to know the origins of the  acronym SWoT, we gather the term was likely introduced by Slavit {\em et al.} in their 2019 paper~\cite{slavit2019stem} offering theoretical perspectives on STEM \textit{Ways of Thinking} (SWoT). Given that most of our research is situated within the context of an introductory physics engineering design (ED)-based laboratory, we found their discussion on viewing SWoT through a disciplinary lens particularly relevant. Based on their empirical study involving K-12 students engaged in an engineering design-based activity, they claim that the notion of SWoT is discipline-specific and influenced by the nature and goals of the design task, the scaffolds provided, and other factors. Notably, they observe that:~\textit{``not much research exists on STEM ways of thinking''}~\cite[p.797]{slavit2019stem}. 

\subsection{WoT Frameworks - Recent Developments}\label{sec:II.D}
In what appears to be a phase of significant development from 2021 to 2023, we identified three studies which saw a movement towards \textit{characterizing} WoT in contexts which have some commonalities with ours. 

Dalal {\em et al.}~\cite{dalal2021developing}, in 2021, proposed their~\textit{Framework for Applying Ways of Thinking in Engineering Education Research} (FAWTEER) tailored to Engineering Education Research (EER) contexts by identifying four ways of thinking: futures, values, systems, and strategic. With a particular focus on characterizing WoT, they assert FAWTEER can build capacity for researchers, and that it is well-suited for collaborative studies. Notably, they also acknowledge that~\textit{``there could be other ways of thinking''}~\cite[p.119]{dalal2021developing} outside of FAWTEER, even within EER contexts, thereby signaling significant opportunities for the development of new WoT frameworks.

In 2022, Slavit {\em et al.}~\cite{slavit2021student, slavit2022analytic} presented an analytic framework for understanding students' thinking in STEM environments, based on their previous studies on how students engage in claim, evidence, and reasoning. We were particularly impressed with their detailed descriptions on how they developed the coding schema for their studies and their justifications for coding decisions. They emphasize the potential of SWoT frameworks as valuable tools for educators, capable of capturing \textit{``transdisciplinary thinking''}~\cite[p.145]{slavit2022analytic} and fostering greater integration across various disciplinary content areas. Most importantly, they indicate a scope for including \textit{computational thinking}~\cite[p.134]{slavit2022analytic} into SWoT frameworks. 

The following year, adding variety to the evolving discourse on SWoT, English~\cite{english2023ways} introduced a novel SWoT framework in the context of~\textit{Mathematical and STEM based problem solving}. With critical thinking, systems thinking, and design-based thinking as its core components, this framework posits that designing learning activities which focus on these facets can collectively enhance students' abilities to tackle complex, ill-defined problems effectively. 

To summarize this sub-section, we emphasize the key commonalities and differences between the reviewed frameworks and our own context. All three frameworks — due to Dalal {\em et al.}, Slavit {\em et al.}, and English—offer valuable perspectives on interdisciplinary and transdisciplinary thinking. However, they are situated in contexts that differ from our focus on Engineering Design (ED)-based physics problems. Specifically, our interest extends to capturing the interplay of design, physics, mathematics, and computational thinking (CT). Notably, although Slavit {\em et al.} suggest a potential for integrating CT, they do not explore it in depth, leaving a critical gap for frameworks addressing integrated STEM education. 

\subsection{Why a new SWoT Framework?}\label{sec:II.E}

\begin{figure*}[!htbp]
\caption{The proposed Ways of Thinking for Engineering Design-based Physics (WoT4EDP) framework. The double-headed arrows illustrate the interconnectedness of each element with the other four \textit{Ways of Thinking}. A non-exhaustive list of pedagogical goals that the framework can support is presented. The concept of \textit{Learning Innovation} is adapted from English~\cite{english2023ways}.}
\fbox{\includegraphics[width=0.96\linewidth]{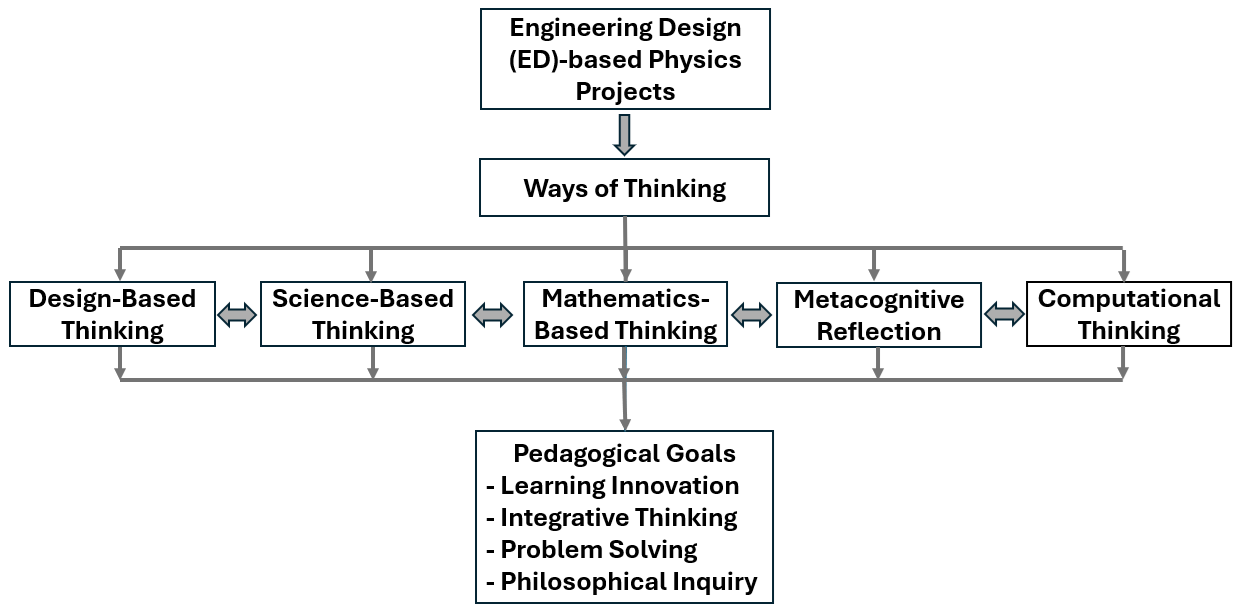}}
\label{fig:WoT_framework}
\end{figure*}

In 2022 the lead author of the current study embarked on an exploratory study~\cite{subramaniam2023narst} into what we termed as the \textit{``Types of Thinking''} that students engage in within an engineering design (ED)-based laboratory task. An ED problem~\cite{capobianco2018characterizing} is client-driven and goal-oriented, set in an authentic context to engage students. It includes constraints such as cost, time, materials, familiar resources, and tools. Given that this is a physics course, we structured the problem to facilitate the application of physics concepts. The solution is a tangible product or process. Another important characteristic of ED problems is that they have multiple solution paths, requiring teamwork to creatively balance constraints and resources~\cite[p.60, 61]{capobianco2013shedding}. 

In 2023, we focused on characterizing the design-science gap in students' thinking within an ED task~\cite{ravi_perc_2023}. Building on this study, in 2024, inspired by ongoing emphasis on SWoT, we transitioned to using the term \textit{STEM Ways of Thinking}~\cite{ravi_perc_2024} and examined how student groups' thinking compared when solving both instructor-assigned and student-generated ED problems. Our recent paper~\cite{ravi_prper_2024} presents an in-depth investigation of the design-science connections in students' thinking while solving an instructor-assigned ED problem (see Figure~\ref{ED_problem_01}) in a physics laboratory. To guide our analysis, we developed and applied a rigorous analytic framework, integrating qualitative research approaches such as traditional coding, thematic analysis, and thick description. Given the lack of universally accepted definitions of \textit{design thinking} and \textit{science thinking}, a key objective was to characterize these constructs within our context. Having established a detailed characterization of students' SWoT in ED-based tasks, we saw an opportunity to contribute to the broader discourse on SWoT frameworks, leading to the development of the WoT4EDP framework.

At this juncture, we confronted a critical question: Is the development of a new framework truly necessary, or would modifying existing ones suffice? As we explain in subsequent sections, the frameworks proposed by Dalal {\em et al.} and English were originally designed for contexts that differ significantly from ours. The elements of Dalal {\em et al.}'s and English's frameworks, in our view, do not always directly apply to physics classrooms, though valuable connections can be made depending on local contexts. Our aim is to propose a framework closely aligned with physics courses that include ED-based experiences while maintaining adaptability for broader application in physics education. Furthermore, we sought to ground our framework in reform documents~\cite{AAPT_2023_lab_reco, NGSS2013, NRC2012, nrc_dber_2012, math_cupm_2015, nrc_CT_2010} and research articles specific to Physics Education Research (PER), which may not have been a primary consideration of Dalal {\em et al.} and English. Notably, neither Dalal {\em et al.} nor English provide detailed, practical guidance on how to apply their models to actual student data, at least at the time of submission of this manuscript. Additionally, neither incorporates computational thinking—a key aspect in our pedagogical interventions. We also emphasize the interconnectedness of the five elements within our framework, a dimension that may not be as thoroughly explored in their works. While we acknowledge the influence of these existing frameworks, our approach extends beyond merely extending them.

A primary motivator for the development of a new framework is the distinct context of our study: an engineering design task embedded within an introductory, calculus-based physics laboratory for engineering students. In this setting, students naturally engage in multiple modes of thinking, such as design thinking and science thinking, which intersect in complex ways. While existing frameworks offer valuable perspectives, our experience reveals the need for a model that captures the nuanced interplay of these thinking processes within the integrated context of engineering design and physics problem-solving. We aimed to provide a practical tool which physics educators can employ for their teaching and which students can leverage to direct their learning. In this paper, we will clearly define each element of our framework, grounded in reform documents that are relevant to physics education. The application of our framework will be addressed in a follow-up paper.

Finally, we recognize and acknowledge an overlap of our framework with contemporary frameworks, given that ours is an integrated STEM context. In particular, we will discuss how WoT4EDP compares with the frameworks of Dalal {\em et al.} and English in extensive detail in the forthcoming sections.  Ultimately, we contend that WoT4EDP represents a simple yet practical framework for integrated STEM physics educators. Building on prior research, we present a refined SWoT framework, highlighting its features and practical applications for physics educators.  

\subsection{Guiding Perspectives}\label{sec:II.F}

In addressing the question, \textit{``Why Are You Doing This Study?''}~\cite[ch.2]{maxwell2013qualitative}, we were guided by Maxwell's \textit{Qualitative Research Design} (2013). We strove earnestly to align our study with the eight key markers of quality in qualitative research outlined by Tracy: (a) worthy topic, (b) rich rigor, (c) sincerity, (d) credibility, (e) resonance, (f) significant contribution, (g) ethics, and (h) meaningful coherence~\cite[p.840]{tracy2010qualitative}. To ensure \textit{reliability} and \textit{validity} (some researchers suggest the use of equivalent terms such as \textit{trustworthiness} and \textit{dependability})~\cite{ozkan2023validity, tracy2010qualitative, golafshani2003understanding, lincoln1985naturalistic} of our framework development process, we anchored our approach in widely recognized reform documents published by organizations such as the National Research Council (NRC), the American Association of Physics Teachers (AAPT), the Mathematical Association of America (MAA), the College Board, and others, along with contemporary research articles. These sources provided a solid foundation for identifying and defining the five elements of our framework, shaping its conceptualization in a manner relevant to Physics Education and our specific educational context~\cite[ch.3]{maxwell2013qualitative}.

Regarding the definitions of the five elements—design-based, science-based, mathematics-based, metacognitive reflection, and computational thinking—we recognize that there is no universally agreed-upon definition for such broad and complex concepts. In this context, we found reassurance in the philosophical view that \textit{``there is no possibility of attaining a single, 'correct' understanding of the world''}~\cite{kiegelmann2002role}. In her evocative essay \textit{The Role of the Researcher in Qualitative Psychology} Kiegelmann argues that researcher's perspective has an important influence on any qualitative study. She aptly quotes Putnam's succinct words that there is no ultimate \textit{``God's eye view''}~\cite[p.15]{kiegelmann2002role}. (Kiegelmann, 2002, p. 15). Accordingly, we advocate for a flexible approach that allows these concepts to be adapted to specific educational contexts. These elements are conceptual constructs rather than fixed, measurable entities, and their interpretation may vary depending on the learning environment and objectives. This inherent flexibility enhances the framework’s capacity to address the complexities of real-world educational settings.

For this study, we received guidance from Briedenhann and Butts’ article, \textit{Utilization-Focused Evaluation}~\cite{briedenhann2005utilization}, which identifies three stages of evaluation, namely: \textit{Knowledge Construction}, \textit{Values in Evaluation}, and \textit{Evaluation Practice}.  \textit{Knowledge Construction} enabled us to examine the key concepts within each WoT framework and their potential contributions to student learning and problem-solving. \textit{Values in Evaluation} prompted reflection on the pedagogical goals and disciplinary priorities we aim to emphasize. Finally, \textit{Evaluation Practice} encouraged us to carefully consider the methods and tools necessary within the context of our framework to effectively achieve our pedagogical objectives.

Consistent with our perspective that there can be no one-size-fits-all WoT framework - particularly in educational contexts - and that human understanding is enriched by having diverse viewpoints, we found Weiss' approach of viewing \textit{``evaluation as enlightenment''}~\cite[p.223]{briedenhann2005utilization} to be particularly valuable. In light of the complexities inherent in educational settings, we also resonate with Scriven's assertion that \textit{``the most complete picture of reality can be constructed through the use of multiple, different perspectives''}~\cite[p.222]{briedenhann2005utilization}.

For the comparison of frameworks, we selected the works of English (2023)~\cite{english2023ways} and Dalal et al. (2021)~\cite{dalal2021developing}. To our knowledge, these are the only available frameworks that have some connection with our context. Additionally, both frameworks provided valuable insights that enriched our understanding of the elements of our framework, making their inclusion both relevant and constructive.

\section{Presenting the W\texorpdfstring{\lowercase{o}}{o}T4EDP Framework}\label{sec:III}

In this section, we address our first research goal \hyperref[RG1]{RG1} by presenting the WoT4EDP framework for Engineering Design (ED)-based projects in introductory physics. This section is organized as follows: Section~\ref{sec:III.A} sets the stage by providing the context for our studies. Establishing this context is crucial as it situates the framework within its educational and disciplinary setting, enhancing the reader’s understanding of the specific needs, challenges, and considerations that informed its development. This not only clarifies our approach but also aids other practitioners in interpreting and adapting the framework for their own educational contexts.

Section~\ref{sec:III.B} details our approach to integrating ED into our physics laboratory course. This section is vital because it offers a concrete understanding of how ED can be incorporated into instructional settings, providing practical insights and actionable steps for educators seeking to implement similar strategies. Section~\ref{sec:III.C} reviews our prior work, that was instrumental in shaping the current framework. Understanding the evolution of our thinking establishes the rationale behind the framework’s components and illustrates the iterative nature of our research process, guiding practitioners in refining their own methodologies.

Section~\ref{sec:III.D} provides a comprehensive breakdown of the five core elements of the framework and Section~\ref{sec:III.E} delves into a non-exhaustive list of pedagogical goals which our framework can facilitate.. Section~\ref{sec:III.F} addresses the limitations of our framework, and finally, Section~\ref{sec:III.G} discusses the broader implications of adopting a Ways of Thinking framework in educational practice.

\renewcommand{\arraystretch}{1}
\begin{table*}[!htbp]
\begin{center}
\captionsetup{justification=raggedright} 
\caption{A list of earlier qualitative studies that contributed to the development of the WoT4EDP framework, all conducted within an introductory physics laboratory course spanning 13 to 14 weeks in a semester. Note: The lead author was a co-author in the Fall 2022 study~\cite{lopez2023promoting}.}

\begin{ruledtabular}
\label{tab:data_earlier_studies} 
\begin{tabular}{p{0.15\linewidth} p{0.25\linewidth} p{0.55\linewidth}}
\textbf{Study situated in} & \textbf{Data}  & \textbf{Research Goals}\\
\hline
Fall 2021: \newline{week 08}   & Transcripts of group discussions within 14 student groups in response to prompts.      & To investigate how students articulated their thinking and engaged in evidence-based reasoning as they described the strategies with which they approached a prescribed ED problem in the lab.~\cite{subramaniam2023narst}\\
Fall 2021: \newline{week 10}   & Final written reports of 27 student-groups. & To explore the level of completion and correctness to which student-groups apply physics concepts to their ED challenge. To investigate evidence for design-science gap.~\cite{ravi_perc_2023}\\
Fall 2022: \newline{weeks 02-06}    & Flowcharts within the written reports of 6 student-groups. & To explore the extent to which engaging students in integrated engineering design and physics labs may impact
their development of computational thinking.~\cite{lopez2023promoting}\\
Spring 2023: \newline{weeks 06 \&\ 14}    &  Transcripts of group discussions within 14 student groups in response to prompts.  & To explore how student-groups’ ways of thinking compare while engaging in two multi-week ED tasks- an \textit{instructor-assigned} ED problem and a \textit{student-generated} ED problem.~\cite{ravi_perc_2024}\\
Fall 2021: \newline{weeks 06 - 10}   & Transcripts of group discussions and final written reports of 14 student-groups & To characterize student-groups' \textit{design thinking} and \textit{science thinking}. To explore the connections between student-groups' design and science thinking. To investigate the influence of scaffolds on student-groups' design and science thinking~\cite{ravi_prper_2024}.\\

\end{tabular}
\end{ruledtabular}
\end{center}
\vspace{-0.5cm}
\end{table*}
\subsection{Context}\label{sec:III.A}

This study, along with our previous research, is set in a large-enrollment, first-semester calculus-based physics course at a major U.S. Midwestern land-grant university. The course typically enrolls about 1,100+ students in the fall semester and 1,400+ students in the spring semester. A vast majority (over 80\%) of the students in this course aspire to be future engineers, while a small fraction are science majors. The course adopts a principle-based approach~\cite{chabay2015matter} such that the content is divided into three units each focused on fundamental physics principles: momentum, energy, and angular momentum. The weekly schedule includes two 50-minute lectures, one 110-minute laboratory, and one 50-minute recitation focused on problem-solving. As part of reforms, ED was integrated into laboratory component of the course in 2019 (see Section~\ref{sec:III.B}). 

Since Fall 2022, we have incorporated computational thinking (CT) tasks into the physics laboratory. Using Jupyter Notebooks~\cite{jupyter} hosted on Google Colab~\cite{googlecolab}, students created flowcharts~\cite{lopez2023promoting} to outline their algorithms and/or coded in Python~\cite{Bralin2023}. These Notebooks, available at no cost to anyone with a Google account, integrate executable code and rich text, making them ideal for running Python scripts, typing equations in LaTeX, uploading images and CSV files, and conducting data analysis. Students were not required to have prior programming experience, as the computational tasks emphasized the application of physics concepts through the editing and modification of existing Python code related to the lab's physical principles. Carefully designed scaffolds supported students in learning programming within the context of their laboratory experiments. The physics tasks included hands-on, inquiry-based experiments using PASCO equipment~\cite{pasco_me5300} and software sensors, with data collection and analysis performed using PASCO Capstone™ software~\cite{pasco_capstone}.

\begin{figure}[!tb]
\caption{ED problem (reproduced from~\cite{ravi_prper_2024}) provided to students in Fall 2021. Wikipedia images were provided, but are not reproduced here. Students tackled this problem alongside scaffolded physics experiments in the laboratory, applying relevant fundamental physics principles, namely: (linear) momentum, energy, and angular momentum~\cite{chabay2015matter}. See~\cite{ravi_prper_2024} and Table~\ref{tab:data_earlier_studies} for details.}
\label{ED_problem_01}
\centering
\begin{tcolorbox}
\justify{``Pristine natural habitats of endangered species such as the gorillas in the Congo River basin are becoming increasingly rare. Today, these habitats and the endangered species that inhabit them need to be not only protected but even sustained by humans. As a member of a team of engineers volunteering for a non-profit organization, you are asked to design a system that can launch a payload of food to an island in the Congo River and land it safely for the gorillas. Each payload is about 50 kg, and it must be delivered to a habitat area located on an island in the Congo River that is about 150 m away from the riverbank. To avoid contributing to global warming, the client wants you to use a means that would minimize the carbon footprint of the delivery. Furthermore, the client also wants to ensure that the habitat remains pristine, so that neither humans nor a robotic machine must disturb the flora and fauna of the habitat while delivering the food.''}   
\end{tcolorbox}
\end{figure}
\subsection{Engineering Design based Labs}\label{sec:III.B}

In parallel with the hands-on and simulation-based lab tasks, student groups of two or three engaged with an ED problem. Seven essential characteristics of ED, as outlined by Capobianco {\em et al.}~\cite{capobianco2013shedding}, guided the development of our ED problem: client-driven and goal-oriented, authentic context, constraints, use of familiar materials, resources, and tools, solution is an artifact or process, multiplicity of solutions, and teamwork. In the context of our study, we adopt the definition of ED as~\textit{``a recursive activity that results in artifacts $-$ physical or virtual $-$ as well as processes''}~\cite[p.348]{capobianco2018characterizing}. This includes stages such as: problem scoping and information gathering, idea generation, project realization, communication and documentation of performance results, and optimization. Similar terms such as \textit{``ill-structured''}~\cite[p.444]{dringenberg2018experiences}, \textit{``wicked''}~\cite{farrell2013design}, and \textit{``ill-defined''}~\cite[p.226]{cross1982designerly} problems also find mention in the literature, and, for our purposes, we shall not delve into any subtle differences that may exist among them. One caveat is that our students did not develop an actual physical setup due to time and resource constraints - typical aspects which may be obvious to those who teach large enrollment courses. We, therefore, focused on how students engaged in the ED process to progress to the solution. 

Two of the ED-based problems that were assigned to students in the Fall 2021 and Fall 2022 semesters can be seen in Figures~\ref{ED_problem_01} and~\ref{ED_problem_02}. Students collaboratively worked in teams to first identify the overall context of the problem and generated possible ideas  using what they knew about the problem as well as using relevant physics knowledge. Even as the teams completed the scaffolded laboratory activities, in parallel, they developed a plan, applied their scientific knowledge, exchanged ideas with other teams, received feedback from the graduate teaching assistants (GTAs), made iterations, refined their solution approaches, and presented their final solution in their lab report. The lead author was a GTA in all the studies (see Table~\ref{tab:data_earlier_studies}).

\begin{figure}[!tb]
\caption{A slightly edited version (reproduced from~\cite{lopez2023promoting}) of the ED problem given to students in Fall 2022. A related YouTube video was shared. The workstation layout-diagram is not reproduced here. Students tackled this problem alongside scaffolded physics experiments in the laboratory, applying relevant physics concepts such as momentum, force, vectors, etc. See~\cite{lopez2023promoting} and Table~\ref{tab:data_earlier_studies} for details.}
\label{ED_problem_02}
\centering
\begin{tcolorbox}
\justify{``A shipping company aims to improve efficiency using modern technology, specifically by automating package handling within its facilities. They plan to implement Automated Guided Vehicles (AGVs) in their new warehouse, which will transport items to workstations where human workers prepare them for shipment. The company has requested your team to develop an algorithm to remotely control the AGVs. This algorithm, along with a flowchart and written description, will be reviewed and then implemented in Python by expert programmers. In a scenario where only one AGV operates across 18 workstations, transport time is constant but varies based on the specific destination. To optimize item allocation and minimize time, the AGV should deliver the first load to the closest station, the second load to the next closest, and so on, ensuring workstations are assigned in proximity to reduce travel distance. Each work station has an area of $0.75 \times 0.75$ m\(^2\). The distance between two neighboring rows of work stations is slightly larger than 0.75 m, allowing AGVs to fit perfectly in between and underneath a work station. However, when carrying a load, the AGV cannot pass through a work station.
The parameters of the AGV robot are as follows: Mass: 145 kg; Dimensions: 75 cm $\times$ 60 cm $\times$ 30 cm; Maximum load: 340 kg; Maximum momentum: 220 kg$\cdot$m/s.''}   
\end{tcolorbox}
\end{figure}

\subsection{The Road to WoT4EDP}\label{sec:III.C}

In this section, we outline the process by which we identified five key \textit{Ways of Thinking} (WoT) through the qualitative analysis of student artifacts from Engineering Design-based physics projects, as detailed in our earlier studies (see Table~\ref{tab:data_earlier_studies}). 

In our initial study~\cite{subramaniam2023narst} we identified four distinct \textit{Types of Thinking}: design, science, mathematical, and metacognitive, using qualitative coding of student work. Building on this foundation, we~\cite{lopez2023promoting} incorporated computational thinking, leveraging student data from a similar ED-based laboratory sequence to further refine the thinking framework. Further,~\cite{ravi_perc_2023},  we investigated the \textit{design–science gap} observed in student projects, providing insight into how these modes of thinking interact. Inspired by ongoing discourse on thinking frameworks~\cite{dalal2021developing, slavit2022analytic, slavit2021student, english2023ways}, we adopted the term \textit{Ways of Thinking} in our next study~\cite{ravi_perc_2024}, aligning our framework with contemporary educational research. These efforts culminated in our recent paper on \textit{design–science connection}~\cite{ravi_prper_2024}, where an elaborate coding scheme was developed to characterize and contextualize student thinking in ED-based physics projects.

Our analysis revealed how students navigated real-world design challenges by combining design thinking, physics principles, reflective thinking, and mathematical reasoning. Briefly, our previous studies focused on understanding how students apply design and physics thinking in tandem, highlighting the importance of interdisciplinary problem-solving in STEM education. The context for this research stems from a growing recognition of the need for students to connect scientific knowledge with engineering practices~\cite{chao2017bridging, NGSS2013, nrc_dber_2012, chase2019learning, kolod_punt_2003}. Key findings from our earlier work showed how students engaged in iterative design processes, applied physics concepts to refine and justify their solutions.

In our studies, we utilized qualitative methods, including coding and thematic analysis, to examine student artifacts such as transcripts of group discussions and written reports (see Table~\ref{tab:data_earlier_studies}). To analyze our data, we developed an inductive coding scheme guided by the Gioia framework~\cite{gioia2013seeking}, integrating thematic analysis~\cite{braun2006using} informed by the MIRACLE framework~\cite{younas2023proposing}. This approach enabled us to provide \textit{thick descriptions}~\cite{stahl2020expanding, brink1987reliability} of our findings.

Qualitative analysis is particularly suited to this work, as it allows for a nuanced exploration of the rich, contextualized details of student thinking. By identifying patterns and themes within these artifacts, we uncovered how student groups' thinking evolved while engaging in design-based tasks. Our inductive coding approach facilitated the organic emergence of themes such as design thinking, physics thinking, and the design-science connection from the data, offering deeper insights into the complexities of students' thought processes. These insights informed the development of our new framework, which captures the multifaceted nature of STEM thinking by integrating design, physics, and problem-solving. This framework not only reflects students’ activities within their projects but also serves as a tool for educators to better understand and foster the development of key STEM thinking skills.

Building on our studies and the growing emphasis on thinking frameworks, we now propose our own Ways of Thinking for Engineering Design-based Physics (WoT4EDP) framework. In this framework, we identify five key elements that can guide students in developing effective solutions to engineering design problems. We believe that an intentional focus on these five aspects can be instrumental in achieving our pedagogical goals (see Figure~\ref{fig:WoT_framework}). These objectives, while not exhaustive, include fostering integrative thinking, enhancing problem-solving skills, and promoting philosophical inquiry. Additionally, we incorporate English's concept of \textit{Learning Innovation}~\cite{english2023ways} as a valuable pedagogical goal.

\subsection{The Five WoT4EDP Elements}\label{sec:III.D}

This section presents the five key elements of our Ways of Thinking framework. Each element addresses a critical aspect of student learning in the context of Engineering Design (ED) problems. Section~\ref{sec:III.D(i)} explores Design-Based Thinking, while Section~\ref{sec:III.D(ii)} focuses on Science-Based Thinking. Section~\ref{sec:III.D(iii)} delves into Mathematics-Based Thinking, followed by Section~\ref{sec:III.D(iv)}, which discusses Metacognitive Reflection. Finally, Section~\ref{sec:III.D(v)} examines Computational Thinking. 

By engaging students in this comprehensive and integrated approach, we aim to achieve our pedagogical goals (elaborated in  Section~\ref{sec:III.E}) while cultivating their 21st-century skills~\cite{jang2016identifying, nrc_21st_century}, including critical thinking, creativity, collaboration, and communication. These skills are essential not only for effective problem-solving but also for fostering innovative thinking and adeptly navigating future challenges in scientific inquiry and engineering design.

\renewcommand{\arraystretch}{1}
\begin{table*}[!htbp]
\begin{center}
\captionsetup{justification=raggedright} 
\captionsetup{width=\textwidth}
\caption{Table (adapted from Dalal {\em et al.}~\cite{dalal2021developing}) presenting the five key elements of the WoT4EDP framework, including a non-exhaustive list of Concepts and Abilities developed, and suggested Enhancement Approaches. Abbreviations: DBT—Design-Based Thinking (or Design Thinking); SBT—Science-Based Thinking (or Science Thinking); MBT—Mathematics-Based Thinking (or Mathematical Thinking); MER—Metacognitive Reflection; CT—Computational Thinking.}

\resizebox{\textwidth}{!}{%
\begin{ruledtabular}
\label{tab:wot_concepts} 
\begin{tabular}{p{0.09\linewidth} p{0.22\linewidth} p{0.33\linewidth}p{0.32\linewidth}}
\textbf{WoT4EDP} & \textbf{Concepts} & \textbf{Abilities developed} & \textbf{Enhancement approaches}\\
\hline
DBT   & Economic aspects, safety aspects, criteria and constraints, identify stakeholders, contraption dimensions, contraption mechanism, design limitations, feasibility study.~\cite{capobianco2018characterizing, ravi_prper_2024}  & Design thinking helps students develop problem-solving skills, creativity, empathy, collaboration, iterative thinking, and multiple perspectives.~\cite{owen2007design}   & Incorporate real-world, open-ended problems to encourage creative and user-centered solutions. Use iterative prototyping and reflective feedback sessions to help students refine ideas and increase empathy. Collaborate across disciplines to stimulate innovation. A focus on failure as a learning opportunity strengthens resilience. The `design thinking competency model'~\cite[p.342]{razzouk2012design} may be used as a guide. \\
SBT   & Use of physics concepts, ideas, and vocabulary. Use of fundamental principles such as momentum, energy, and angular momentum. Detailing assumptions and approximations made.~\cite{collegeboard2009science, NGSS2013, AAPT_2023_lab_reco, ravi_prper_2024} & Science thinking fosters the ability to ask questions, formulate hypotheses, and design experiments to test those hypotheses. Students learn to apply scientific principles to solve problems. It teaches students to analyze data and develop evidence-based arguments.~\cite{osborne2010arguing} & Encourage inquiry-based learning and hands-on experimentation. Promote hypothesis generation, testing, and iterative feedback, focusing on evidence-based reasoning. Incorporating argumentation and critical evaluation of data helps students make more informed, reflective decisions about scientific concepts.\\
MBT    & Identify variables and constants. Use of basic algebra, trigonometry, geometry etc., proportional reasoning, dimensional analysis, use of advanced tools such as calculus. Qualitative reasoning. Use graphs and tables to depict and identify patterns.~\cite{math_cupm_2015, tiles1984mathematics, ravi_prper_2024}   & Mathematical thinking helps students develop the ability to break problems down into manageable parts and solve them using logic and structured reasoning. Students gain skills in recognizing patterns, relationships, and abstract concepts, which are key to problem-solving in math and STEM-related fields.~\cite{chapman2018mindset, daly2019mathematical} & Use real-world applications and collaborative problem-solving to make abstract concepts concrete. Encourage multiple approaches to solving problems, fostering flexibility and deeper conceptual understanding. Integrate technology, such as graphing calculators or Python, to enhance exploration and visualization of complex mathematical ideas. \\
MER   & Reflecting on their design, discuss the need for iterations, reflecting on various science concepts to be used.~\cite{nrc_meta_2000, dori2018cognition, hacker2003not, ravi_prper_2024} & Metacognitive reflection encourages students to think about how they are learning, make adjustments to their strategies, and evaluate the effectiveness of those strategies. Through reflection, students learn to adapt their thinking and approaches to new problems, enhancing their capacity for lifelong learning.~\cite{PriceMitchell2015, stanton2021fostering} & Encourage self-reflection and awareness of one’s cognitive processes through reflective journaling and self-assessment activities. Teach students to plan, monitor, and evaluate their problem-solving strategies. Peer collaboration and discussion can also help students refine their metacognitive strategies.  \\
CT  & Decomposition, Pattern Recognition, Abstraction, Algorithm Design, Automation, and Debugging.~\cite{shute2017demystifying, lopez2023promoting} & CT fosters an understanding of algorithms, logic, and the breakdown of problems into manageable steps for computational solutions. Students work with data, automate processes, and improve their efficiency in addressing complex tasks, which is crucial for data analysis and modeling. Additionally, CT enhances programming skills through the use of languages such as Python~\cite{grover2013computational}. & Encourage the use of algorithmic problem-solving by tackling real-world coding projects. Emphasize best practices such as code readability, modularity, and efficiency. Engage students in debugging, refactoring, and peer-review sessions to enhance their coding strategies and collaborative thinking. \\

\end{tabular}
\end{ruledtabular}
}
\end{center}
\end{table*}

\subsubsection{Design-Based Thinking}\label{sec:III.D(i)}

Based on our readings~\cite[p.4]{laakso2014promoting}, we gather that the term \textit{design thinking} likely originated with Rowe, who used it as the title of his 1987 book~\cite{rowe1991design}, in which he aimed to provide a structured account of problem-solving processes in design. Since then, however, the meaning of the term design thinking has undergone several transformations and has drawn varied interpretations.

Haasi and Laakso~\cite[p.5]{laakso2014promoting} describe design thinking as comprising three key elements: (i) practices, including human-centered approaches, thinking by doing, visualizing, integrating multiple approaches, and collaboration; (ii) cognitive approaches, such as abductive reasoning, reflective reasoning, a holistic perspective, and integrative thinking; and (iii) a mindset characterized by exploration, tolerance for ambiguity, optimism, and a future-oriented outlook.

Despite its widespread usage, Kimbell~\cite{kimbell2011rethinking}, as recently as 2011, notes with caution that \textit{``just what design thinking is supposed to be is not well understood, either by the public or those who claim to practice it''}. As much as the term is popular, interestingly, it has its fair share of critics. Normann calls it a \textit{``useful myth''} and that it is merely a \textit{``public relations term for good, old fashioned creative thinking''}~\cite{useful_myth}. Nussbaum even calls it a \textit{``failed experiment''}~\cite{nussbaum_dt_failed}. 

While the debate over the usage of design thinking is thought-provoking, we find merit in the use of the term in our contexts. For operational purposes, one may view design thinking as a problem-solving approach that emphasizes creativity, user-centered solutions, and iterative processes. It involves understanding users' needs, brainstorming ideas, prototyping, and testing solutions. Widely applied in fields such as engineering and business, design thinking fosters innovation by blending empathy, experimentation, and practical implementation~\cite{kimbell2011rethinking, razzouk2012design}. In the context of school education, Li {\em et al.} argue that  design thinking should be regarded as a model of thinking to enhance students' learning~\cite[p.97]{li2019design}. They cite several studies showing how students can learn through design based activities and also develop design thinking skills~\cite[p.98]{li2019design}. A valuable resource for educators is the \textit{Educational Designer}~\cite{educationaldesigner}, an international open-access e-journal focused on research-based practices in the design, development, and evaluation of educational materials in mathematics, science, engineering, and technology\cite[p.101]{li2019design}.

With the growing emphasis on design thinking in education~\cite{NGSS2013, itea2007standards, love2016technological}, educators have numerous opportunities to integrate it into their curricula. In a university-level physics course, design-based problems provide a context for students to not only apply theoretical principles but also expand their understanding of science. Iterative engagement in design fosters better solutions and enhances students' scientific knowledge. Beyond academic benefits, it cultivates a mindset for innovation and interdisciplinary work, skills that will be invaluable in their professional lives.

It is to be acknowledged that assessing students' design thinking and tracking its development over time remains a significant challenge~\cite{li2019design}. Our earlier work~\cite{ravi_prper_2024} focused on characterizing students' design thinking (see Table~\ref{tab:data_earlier_studies}) and explored the role of scaffolds~\cite{dasgupta2019improvable} within the context of an Engineering Design (ED) problem. However, more experimental studies are needed to examine how various educational interventions influence the development of students' design thinking over time.

To engage students effectively in design-based thinking (or, design thinking), we found the guidance offered by Capobianco~{\em et al.}~\cite[p.346, 348]{capobianco2018characterizing} particularly useful in our context. Design-based thinking, especially in engineering design-based science learning, is an iterative, problem-solving process that integrates inquiry, reflection, and collaboration. It begins with problem scoping and information gathering, where students identify the problem, key stakeholders (such as users or clients), criteria for success, and constraints. They also consider assumptions, trade-offs, and approximations in order to guide their design decisions. This phase encourages students to explore existing solutions, connect with prior knowledge, and gather any additional information required to address the problem effectively.

During solution formulation, students brainstorm individually and collaboratively, identify necessary materials, determine what metrics to measure, and align their design with relevant scientific principles. The solution production and performance phase focuses on building prototypes or models, testing them, and recording results to evaluate how well the design meets the original criteria and constraints. Students must reason through trade-offs, assessing how modifications may impact performance, and interpret their results using scientific and mathematical concepts.

In the communication and documentation phase, students share their designs, compare performance with other teams, reflect on patterns in their results, and incorporate feedback to refine their solutions. This step fosters flexibility and critical thinking by supporting a diversity of ideas and interpretations. The final optimization phase emphasizes refining solutions based on testing outcomes, retesting designs, and identifying which approaches best address the problem requirements. Throughout this process, students engage in cycles of reflection, iteration, and improvement, balancing trade-offs, constraints, and assumptions.

For example, in tackling the ED problem described in Figure~\ref{ED_problem_01}, students may consider various aspects in developing a mechanism to deliver food safely and sustainably to the gorillas' habitat. They might identify key stakeholders, such as the non-profit organization, the client, and the gorillas, whose well-being depends on the food delivery. Students could explore criteria for success, such as ensuring the 50 kg payload reaches the island 150 meters away without damage, while minimizing the carbon footprint and keeping the habitat undisturbed. They may also define constraints, such as avoiding human or robotic interference with the environment and using sustainable delivery methods. Students might evaluate trade-offs between efficiency, cost, and environmental impact—for instance, determining whether the use of motorized solutions is worth the potential emissions. They could brainstorm various contraptions or mechanisms, including catapults, slingshots, ziplines, or renewable-powered drones, to launch the payload across the river. Additionally, they may decide on key metrics, such as range, accuracy, stability, and landing force, to evaluate their design. Students might also need to make assumptions and approximations about factors such as wind conditions or terrain. Through multiple rounds of testing and iteration, students could refine their ideas, balancing design decisions with the need to protect the habitat and meet the delivery goals~\cite{ravi_prper_2024}.

In our context, we interpret design thinking (see Table~\ref{tab:wot_concepts}) as encompassing how students approach the various aspects of a problem. This includes, but is not restricted to, considering stakeholders, defining metrics, criteria, and constraints, making assumptions and approximations, and weighing trade-offs. It also involves providing detailed descriptions of the physical dimensions of the model or contraption, explaining its working mechanisms, recognizing limitations, justifying iterations, and conducting feasibility studies. However, we do not claim that this list to be exhaustive or rigid. Instead, we believe that educators should retain flexibility in determining which aspects to emphasize and to what extent, based on the specific context, constraints, and pedagogical goals of their learning environment. True to this spirit, in all our studies, our approach has been to adopt an inductive approach to analyzing student data. This will be evident in the second installment of this two-part series. 

\subsubsection{Science-Based Thinking}\label{sec:III.D(ii)}

The \textit{College Board Standards for College Success}~\cite[p.5]{collegeboard2009science} outlines five core science practices that shape scientific inquiry. These include: (i) formulating empirically testable questions and predictions, (ii) collecting relevant data to address these questions, (iii) identifying patterns in data through analysis, (iv) constructing scientific explanations using evidence and knowledge, and (v) applying mathematical reasoning to interpret data and solve problems. These practices emphasize critical thinking and methodological rigor in science education. 

The \textit{Framework for K-12 Science Education}~\cite{NRC2012}, ~\cite[p.xv]{NGSS2013} identifies three essential dimensions for providing a high-quality science education: \textit{Science and Engineering Practice}, which engage students in scientific inquiry and problem-solving; \textit{Crosscutting Concepts}, which include overarching themes such as patterns and systems that apply across disciplines; and \textit{Disciplinary Core Ideas}, which focus on the key ideas in physical, life, earth, space, and other sciences. Together, these dimensions help students not only understand scientific content but also grasp how it is acquired and the connections that exist between various scientific disciplines.

The \textit{Next Generation Science Standards} 2013 document also emphasizes  the importance of \textit{``instructional flexibility''}~\cite[p.xiv]{NGSS2013}, empowering educators to tailor their approaches based on local contexts to enhance student learning. This flexibility is crucial for integrating disciplines such as mathematics and computation, which naturally intersect with scientific inquiry. 

The \textit{AAPT Recommendations for the Undergraduate Physics Laboratory Curriculum}~\cite[p.2,3]{AAPT_2023_lab_reco} report, published by the American Association of Physics Teachers (AAPT), emphasizes promoting physics-based thinking through several key focus areas. (i) \textit{Constructing knowledge}: Students are encouraged to independently collect, analyze, and interpret real data, building confidence in their ability to generate scientific insights. (ii) \textit{Modeling}: Developing abstract representations of physical systems is vital, enabling students to predict outcomes, interpret results, and recognize model limitations. (iii) \textit{Designing Experiments}: Students need to be guided to ask scientific questions, engineer solutions, and troubleshoot, gaining valuable hands-on experience throughout the process. (iv) \textit{Developing Technical and Practical Laboratory Skills}: Proficiency with laboratory equipment and understanding the appropriate tools for different measurements are key priorities. (v) \textit{Analyzing and Visualizing Data}: Students are expected to apply statistical methods, quantify uncertainties, and effectively represent findings through data visualization. (vi) \textit{Communicating Physics}: Students learn to present scientific results in authentic formats, such as reports and presentations, while developing teamwork and ethical communication skills. Together, these focus areas aim to cultivate a comprehensive set of skills essential for scientific inquiry.

While our course is an introductory physics class primarily emphasizing disciplinary knowledge in physics—especially Newtonian Mechanics—we remain open to integrating concepts from other areas of physics, and science in general. Our approach to fostering physics-based thinking not only enhances students' understanding of physical principles but also serves as a guideline for broader scientific thinking across various contexts. This openness allows students to explore interdisciplinary connections and apply scientific reasoning in diverse problem-solving scenarios.

In the ED problem shown in Figure~\ref{ED_problem_01}, students’ science-based thinking (or, science thinking) may include constructing knowledge through the collection and analysis of real data related to delivery mechanisms such as ziplines, drones, or catapults. Alternatively, they may apply their physics knowledge to propose solutions. Each approach may involve distinct scientific concepts or ideas: ziplines might require an understanding of tension and gravitational forces, while drones introduce concepts of aerodynamics and propulsion. Students might also consider applying the momentum principle~\cite{chabay2015matter} to evaluate how the payload’s momentum changes throughout its journey, ensuring stability and precision. When working with catapults, they may analyze projectile motion, accounting for factors such as launch angle, initial velocity, and air drag to achieve accurate delivery. The energy principle~\cite{chabay2015matter} can also come into play, as students explore the conversion of potential energy into kinetic energy, optimizing their designs for maximum range and effectiveness. Rather than unfolding in isolated silos, design thinking and science thinking appear naturally interwoven throughout these activities. This interplay, revealed in our analysis of student data, motivated us to advocate for the usage of \textit{design-science connection} as opposed to \textit{design-science gap} in our recent study~\cite{ravi_prper_2024}.

In addition, students may explore material science by evaluating the properties of materials suitable for their proposed delivery mechanisms. Factors such as weight, strength, flexibility, and environmental impact may guide their material selection. For example, lightweight composites might be chosen for drones to enhance flight efficiency, while durable yet eco-friendly materials could be prioritized for ziplines to minimize ecological disruption. When selecting packaging materials for the food payload, students may investigate biodegradable or sustainable options that would not harm the environment. For parachute designs, they might consider materials and shapes that provide adequate drag while remaining lightweight and easy to deploy, ensuring the safe descent of the payload.

By integrating these considerations, students can apply physics concepts and principles~\cite{ravi_prper_2024} to develop comprehensive and effective solutions for delivering the payload while maintaining the integrity of the gorillas' habitat. Furthermore, they may develop mathematical models to predict payload trajectories or analyze flight paths, helping them understand limitations and inform design iterations—an approach which may involve applying their mathematical and computational skills (see sections~\ref{sec:III.D(iii)} and~\ref{sec:III.D(v)}).

It is important to note that, due to resource constraints, our students do not create actual physical models; instead, they will propose solution approaches or conceptual models. Additionally, they might consider concepts beyond physics, such as environmental science, to assess the impact of their delivery methods on the gorillas' habitat. Finally, students may focus on communicating their findings through reports and presentations, fostering teamwork and ethical discourse. This integration of physics-based thinking, along with environmental and material science considerations, supports technical design aspects while promoting critical reflection and iterative problem-solving essential for delivering the payload and preserving the gorillas' habitat.

In our recent study~\cite{ravi_prper_2024}, we examined the extent and depth of students' use of physics vocabulary and their application of fundamental principles such as momentum, energy, and angular momentum to enhance their design solutions. We also analyzed the assumptions and approximations students made, the rationale behind their statements, and how they integrated mathematics into their scientific reasoning. This multifaceted approach, while not exhaustive or inflexible, provided valuable insights into students' conceptual understanding and their application of physics in problem-solving contexts (see Table~\ref{tab:wot_concepts}).

\subsubsection{Mathematics-Based Thinking}\label{sec:III.D(iii)}

The \textit{Committee on the Undergraduate Program in Mathematics} in its 2015 report, emphasizes the need to foster a \textit{``mathematical habits of mind''}~\cite[p.10,11]{math_cupm_2015} by developing students' cognitive abilities and content knowledge. For cognitive goals, the committee recommends: (i) developing students' thinking and communication skills to reason logically and express mathematical ideas clearly, (ii) linking mathematical theory with interdisciplinary real-world applications, (iii) using technological tools effectively for problem-solving and exploration, and (iv) fostering mathematical independence through open-ended inquiry. As for the content, the committee recommends undergraduate programs should cover core areas such as calculus, linear algebra, data analysis, computing, and mathematical modeling. A similar perspective is articulated in the 2013 report by the \textit{Committee on the Mathematical Sciences in 2025}~\cite{nrc_math_2025}, which emphasizes the need to adapt undergraduate education to meet future demands in research, industry, and interdisciplinary applications.

Boaler highlights the importance of promoting a \textit{``growth mindset''} in students by exposing them to open-ended tasks that involve mathematics. This approach can lead to higher engagement and achievement, contrasting with closed questions where frequent wrong answers may discourage effort~\cite{boaler2013ability}. Engagement and motivation in solving mathematical problems can be significantly increased through open-ended tasks that allow for diverse pathways and approaches~\cite{daly2019mathematical}.

In our context, mathematics serves as an indispensable tool for solving engineering design-based problems. This perspective is further reinforced by Tiles’ assertion that mathematics \textit{``is not just a language which can be translated into and out of, rather it plays an essential role in the articulation of scientific concepts''}~\cite[p.3]{tiles1984mathematics}.

As an example, for the engineering design problem in Figure~\ref{ED_problem_02}, students may explore various aspects of mathematics-based thinking (or, mathematical thinking) that are intricately connected to aspects of design and science~\cite{ravi_prper_2024} in the implementation of Automated Guided Vehicles (AGVs) within a warehouse setting. They might start by identifying relevant variables and constants, such as the distances between workstations, the mass of the AGV, its dimensions, and its maximum load capacity, along with constants such as the constant transport time across workstations and the area of each workstation. To optimize the delivery process, students may employ basic algebra to calculate travel distances and determine the optimal sequence for the AGV's deliveries. Geometry and trigonometry may also be essential as they model the warehouse layout and derive necessary distances and angles. Dimensional analysis could ensure consistency in units, while advanced mathematical tools such as calculus might be applied to analyze rates of change or optimize paths based on real-time data.

As may be evident, these mathematical considerations are not isolated; they are closely intertwined with design-based and science-based thinking. Computational thinking (see Section~\ref{sec:III.D(v)}) also plays a key role here. For instance, as students write Python code to create graphs and tables that help visualize their findings, they can identify patterns in travel times and loads, which directly inform their algorithm development for item allocation. Implementing these algorithms in Python allows for simulation and real-time control of the AGVs, merging mathematical modeling with practical design solutions. Additionally, outlining the algorithm steps using flowcharts~\cite{lopez2023promoting} clarifies the decision-making processes governing AGV operations, further blending mathematical reasoning with design principles. By recognizing that mathematics-based thinking is deeply related to design and scientific inquiry, students can effectively address the engineering design problem and enhance the efficiency of the shipping company’s package handling processes.

In our previous studies, we examined how students' mathematical thinking (see Table~\ref{tab:wot_concepts}) influenced their approach to design problems. Specifically, we analyzed how students identified variables and constants, utilized basic algebra, trigonometry, and geometry, and employed proportional reasoning and dimensional analysis. Additionally, we explored their use of advanced tools, such as calculus, and qualitative reasoning. We also investigated how students utilized graphs and tables to depict and identify patterns. It is important to note that this list is far from exhaustive. Our approach has been to remain responsive to the data and observe the patterns that emerge from it. By continually analyzing student interactions and written work, we aim to refine our understanding of how mathematical thinking contributes to solving ED-based physics problems. 

\subsubsection{Metacognitive Reflection}\label{sec:III.D(iv)}

The National Research Council's \textit{How People Learn: Brain, Mind,
Experience, and School: Expanded Edition (2000)}~\cite{nrc_meta_2000} serves as a foundational document that underscores the significance of metacognition in student learning. The report argues that adopting a metacognitive approach to instruction \textit{``can help students
learn to take control of their own learning by defining learning goals
and monitoring their progress in achieving them''}~\cite[p.18]{nrc_meta_2000}. This perspective is built on the understanding that metacognitive thinking encourages students to engage in an \textit{``internal conversation''}~\cite[p.18]{nrc_meta_2000} that fosters self-regulation and self-directed learning. Furthermore, it posits that metacognition can empower students to \textit{``teach themselves''}~\cite[p.50]{nrc_meta_2000}. Consequently, the report advocates for the integration of metacognitive skills into various subject areas within the curriculum.

In the context of this report, the lead author `reflected' on the definition of metacognition. Initially, he perceived it simply as \textit{``reflecting, or thinking about thinking''}. However, he realized this may be a naive view considering Crawford and Capps' assertion that \textit{``Reflection and metacognition are both thinking processes. Yet, reflection and metacognition differ in some ways''}~\cite[p.16]{dori2018cognition}. Our aim in this article is not to engage in a debate regarding the precise definition of metacognition. Rather, it is perhaps prudent to align with Livingston's assertion that \textit{``In actuality, defining metacognition is not that simple''}~\cite[p.2]{livingston2003metacognition}. Given the complexities surrounding the term, we find it beneficial to draw on the fine example provided by Kohen and Kramarski, who note that \textit{``presenting data in a graph is a cognitive function, whereas reflecting on the answer and realizing that the graph fits the givens are part of the metacognitive process''}~\cite[p.280]{dori2018cognition}. For the records, we consider metacognition as a \textit{``deliberate, planful, and goal-oriented mental process, using higher level thinking skills applied to one’s thoughts and experiences''}~\cite[p.17]{dori2018cognition}.

A compelling reason for incorporating metacognitive reflection as a critical way of thinking within our framework is best articulated in Hacker and Dunlosky's thoughtful essay, \textit{Not All Metacognition Is Created Equal}. They emphasize that students \textit{``should not be instructed to merely use `metacognition' but instead should be directed toward specific, higher-level analyses in which they explain how they are solving the problem''}~\cite[p.79]{hacker2003not}. This insistence on directing students toward higher-level analysis highlights the need for metacognitive reflection to be not only present, but also be purposeful in guiding students toward more profound understanding and effective problem-solving strategies.

A similar perspective is echoed in the NRC's 2012 report, which highlights how metacognitive approaches in STEM education are embedded in problem-based learning and reflective activities designed to promote critical thinking. While instructors may assume students possess these skills or view them as too advanced for introductory courses, research shows that explicitly teaching metacognitive skills alongside science content enhances student responsibility, problem-solving abilities, and study habits~\cite[p.154]{nrc_dber_2012}.

Within our context of physics education, the studies of Sayre and Irving~\cite{sayre2015brief} further emphasize the importance of metacognitive reflection to promote student learning. Their study identifies a new element in physicists' discourse called \textit{``brief, embedded, spontaneous metacognitive talk (BESM talk)''}. This type of dialogue reflects students' expectations about physics allowing them to identify themselves as physicists or non-physicists. The presence of BESM talk serves as an indicator of students' metacognitive reflections and their alignment with \textit{`Thinking Like a Physicist'} (TLP)'~\cite{van2010documenting, manogue2010upper, sayre2015brief}, highlighting its significance in fostering professional growth within the realm of physics education.

Another recent study by Ulu and Yerdelen-Damar~\cite[p.20, 21]{metacog_ulu_2024} employs a metacognition framework inspired by Brown, Schraw, and Dennison~\cite{brown1987metacognition, schraw1994assessing}. Metacognition refers to understanding and regulating one’s own thinking and is divided into two components: (i) \textit{Knowledge of cognition}—declarative information about one’s cognitive processes—and (ii) \textit{Regulation of cognition}—the ability to plan, monitor, control, and evaluate one’s thinking. Knowledge of cognition consists of three levels: declarative knowledge (facts and strategies), procedural knowledge (how to apply strategies), and conditional knowledge (when and why to apply strategies). Regulation of cognition involves five key skills: planning, managing information, monitoring comprehension, debugging, and evaluating. Although we did not adopt this framework in our earlier studies, it holds promise for future integration into our WoT4EDP framework. Its structured approach to analyzing metacognitive thinking aligns with the reflective practices we aim to foster in design-based activities.

In our earlier studies~\cite{ravi_prper_2024}, we acknowledged the complexities surrounding the precise definition of metacognition but opted for a simplified interpretation, as detailed in Table~\ref{tab:wot_concepts}. Our goal was not to delve into the rigorous distinctions of what strictly constitutes metacognitive thinking but to adopt a more practical approach. We focused on understanding how students reflected on their design decisions, the rationale behind their iterations, and their review of scientific concepts to facilitate progress toward a solution. By encouraging students to articulate their reflections on the necessity for iterations and their integration of various scientific principles, we aimed to highlight the role of metacognitive processes in enhancing design-based learning outcomes.

In summary, metacognitive reflection serves as a vital component of student learning, enabling learners to navigate their educational journeys with greater agency. By integrating metacognitive strategies into various subject areas, educators can foster a more reflective, self-directed approach to learning that aligns with the goals outlined in the NRC report and addresses the complexities inherent in defining and understanding metacognition.

\subsubsection{Computational Thinking}\label{sec:III.D(v)}

Integrating computational practices within STEM classrooms equips learners with an authentic understanding of scientific processes and better prepares them for careers in an increasingly computational world. As computational thinking (CT) becomes essential across STEM disciplines, the demand for computation-integrated physics courses continues to grow, underscoring the notion that computation is synonymous with engaging in STEM. The urgency and significance of incorporating computational thinking (CT) based activities into physics education is powerfully highlighted by Hamerski {\em et al.}~\cite[p.23]{hamerski_CT_2022}, who raise a \textit{``call to action''} in their case study of high school students involved in a computation-integrated physics course. They urge researchers to recognize and respond to the pressing need for integrating computation into educational practices.

Interestingly, while the term computational thinking (CT) is quite commonplace, its meaning has been the subject of ongoing debate, particularly highlighted in the National Research Council's \textit{Report of a Workshop on the Scope and Nature of Computational Thinking} in 2010~\cite{nrc_CT_2010}. Despite extensive discussions, the committee refrained from offering formal recommendations, underscoring the inherent complexity of defining CT. Notably, the focus was on exploring what \textit{``computational thinking for everyone''} might entail. Given the diverse educational contexts in which CT is applied, we adopt a balanced view that a one-size-fits-all definition may neither be practical nor desirable. Instead, embracing multiple perspectives allows for the adaptation of CT definitions to align with specific local contexts, pedagogical goals, available resources, and disciplinary needs.

Wing's influential definition of CT as \textit{``solving problems, designing systems, and understanding human behavior by drawing on concepts fundamental to computer science''}~\cite[p.33]{wing2006computational} demonstrates that the concept extends far beyond computer science alone. This definition also establishes the relevance of CT to STEM fields, particularly in the context of engineering design (ED)-based problems, which focus on developing human-centered, science-based solutions for real-life challenges~\cite{lopez2023promoting}. In our teaching context, we find Berland and Wilensky’s more approachable interpretation—\textit{``the ability to think with the computer-as-tool''}—to be both compelling and simple~\cite[p.630]{berland2015_CT}.

While educators may choose various computational thinking (CT) frameworks~\cite[p.151, 152]{shute2017demystifying} that best suit their contexts, we have opted for Shute {\em et al.}'s framework due to its relevance to our instructional goals. Our students work extensively with Python in the Jupyter Notebook~\cite{jupyter} environment via Google Colab~\cite{googlecolab}, which makes Shute {\em et al.}’s definition of CT particularly relevant, simple, and practical: \textit{``CT as the conceptual foundation required to solve problems effectively and efficiently (i.e., algorithmically, with or without the assistance of computers), with solutions that are reusable in different contexts''}~\cite[p.153]{shute2017demystifying}. This CT framework comprises six key facets: (i) \textit{Decomposition}—breaking a complex system or problem into manageable, functional sub-problems; (ii) \textit{Abstraction}—extracting key elements, including data collection, pattern recognition, and modeling; (iii) \textit{Algorithms}—developing logical, step-by-step instructions for solving problems, often represented through flowcharts and implemented in code (e.g., Python); (iv) \textit{Debugging}—identifying and fixing errors in code or algorithms; (v) \textit{Iteration}—refining the algorithm or code through repeated improvements; and (vi) \textit{Generalization}—applying algorithms, code, or CT skills across different domains, enabling efficient problem-solving~\cite[p.153]{shute2017demystifying}.

Importantly, Shute {\em et al.} emphasize that this framing of CT underpins and is evidenced through a variety of subject areas, \textit{``including Mathematics, Science, and even English Language Arts''}. This perspective reinforces how various ways of thinking—mathematical, scientific, and computational—are closely linked, facilitating a holistic approach to problem-solving in our educational practices~\cite[p.146, 151]{shute2017demystifying}.

In the context of physics education, Gambrell and Brewe~\cite{gambrell_brewe_CT_2024} explored the learning goals for computationally integrated introductory physics classes, which resonate with our ED-based laboratory. They found that while the core goals—learning and applying physical principles—are consistent with traditional physics courses, computational tools (e.g., Python, VPython, spreadsheets) offer new ways to engage with these concepts. Students also develop coding skills, such as loops, commenting, and creating variables, enriching their ability to represent physics. Importantly, they emphasize that assessments should focus on foundational computational skills, presented in a way that fosters student confidence and ensures that CT supports deeper understanding of physics.

In a previous study~\cite{lopez2023promoting}, we applied Shute {\em et al.}'s CT framework to analyze flowcharts from lab reports submitted by student groups in response to a designated ED challenge. In the second installment of the two-part series, we shift our focus to the students' Python code as they addressed the ED problems they developed. Our analysis will begin with an inductive approach to uncover insights from the data, followed by a targeted application of Shute {\em et al.}'s framework. While we emphasize Python and flowcharts in our studies, we acknowledge that CT extends beyond these tools; however, this narrowed focus aligns with our instructional objectives.

\subsection{Pedagogical Goals}\label{sec:III.E}

In this subsection, we outline the pedagogical goals that this framework can effectively foster. While our goals are rooted in our combined experience as educators and researchers, they are also informed by various reform documents. These goals are neither rigid nor exhaustive; rather, we see them as adaptable to the specific needs of educators and researchers. We recognize that these terms may be interpreted differently across contexts, but we view this diversity as a strength rather than a limitation. Following Gee, we conceptualize these terms as \textit{``thinking devices''}~\cite[p.6]{gee2014discourse}, offering flexibility in their interpretation and application.

Though learning physics is not solely about applications~\cite{thring1965nature, weisskopf1965pure,sirnoorkar2020towards,sirnoorkar2016students}, the \textit{Phys21: Preparing Physics Students for 21st Century Careers (2016)} report underscores the importance of connecting physics to real-world applications~\cite[p.26]{phys_21_2016}. This integration, which embodies \textit{Integrative Thinking}, goes beyond connecting sub-fields within physics, extending to interdisciplinary links among physics, math, biology, and beyond, as well as the various ways of thinking outlined in our framework. The report also highlights concerns that current curricula are lacking in key areas such as contextual understanding of core concepts, real-world problem-solving, collaboration, communication, and technology skills, along with lifelong learning and innovation~\cite[p.11]{phys_21_2016}. We believe that a deliberate focus on WoT frameworks—not just our own—can help address these concerns.

\textit{Problem Solving} is a most natural goal for any science educator, yet it is almost taken for granted. The \textit{AAPT Recommendations for the Undergraduate Physics Laboratory Curriculum} emphasize that physics is an approach to problem-solving that relies on direct observation and hands-on experimentation. To succeed in this field, one must integrate a wide range of knowledge and skills, including mathematical, computational, experimental, and practical abilities, while also cultivating a \textit{``thinking like a physicist''} mindset~\cite[p.iii]{AAPT_2023_lab_reco}. A strong grasp of scientific principles is important for students to effectively and efficiently solve problems. As the Next Generation Science Standards (NGSS) 2013 report rightly notes, \textit{``higher-level reasoning and problem-solving practices require a reasonable depth of familiarity with the content of a given scientific topic if students are to engage in them meaningfully''}~\cite[p.15]{NGSS2013}.

Regarding \textit{Learning Innovation}, we are largely guided by English’s work, and we will return to this in more detail in Section~\ref{sec:IV.B(iv)}. While innovation in this context refers to students' learning, we are tempted to extend it as a goal for educators too. We would approach the term \textit{innovation} rather cautiously, and the \textit{SPIN-UP Report} aptly captures our tempered view: true \textit{Eureka} moments are rare in science, and even rarer in education. Progress in teaching and learning is incremental and ongoing, with innovations needing to be adapted to suit local needs~\cite[p.49, 67]{spin_up}.

\textit{Philosophical Inquiry}, though it may seem esoteric to some, is included in our goals to encourage students to think deeply about scientific ideas, not merely as tools for problem-solving, but as pathways to a deeper understanding of nature. To our support, we cite Burgh's emphatic recommendation: 
\begin{quote}
\textit{``Philosophical inquiry as pedagogy must be integrated into the curriculum as well as integrate the curriculum.''}~\cite[p.1057]{burgh2012parallels}.
\end{quote}
Gale pushes this notion further, arguing that \textit{``not only are the domains of philosophical inquiry and physical inquiry co-extensive, but also the results of these two inquiries will tend to be both co-extensive and co-intensive''}. He further speculates that in the future, universities may merge Physics and Philosophy into a single department, perhaps called the \textit{``Department of Theoretical Inquiry''}~\cite{philo_phys_gale}. While we may not be there yet, his reflections inspire us to integrate deeper philosophical thinking into science education.

\subsection{Limitations of our Framework}\label{sec:III.F}
The development of the WoT4EDP framework is context-specific, focusing on the particular engineering design-based physics projects and student groups involved, which may possibly limit its generalizability to other contexts or disciplines without further study. It is important to note that the Ways of Thinking are not mutually exclusive, and overlap is inevitable. In fact, in our recent study we discussed in length how students' thinking does not occur in silos and is far too interconnected in quite unpredictable ways~\cite{ravi_prper_2024}. While the framework was developed through multiple qualitative studies~\cite{subramaniam2023narst, ravi_perc_2023, ravi_prper_2024, ravi_perc_2024}  of student artifacts (see Table~\ref{tab:data_earlier_studies}), it did not account for critical factors such as individual prior knowledge, motivation, communication skills, group composition, role assignment, group dynamics, physical environment, time constraints, cultural context, peer pressure, and confidence levels~\cite{schmidt2000factors}. Furthermore, additional artifacts such as individual reflections, video recordings of group interactions, and assessments of student work could provide deeper insights into the nuances of student thinking. Instructional factors such as the cognitive load imposed by the learning task, materials, and instruction; the scaffolding and feedback provided by the instructional materials, as well as the influence of teaching assistants (TAs), were not examined in our study, even though they likely affect student thinking and outcomes~\cite{huffmyer2019graduate}. Future work could address these factors to refine and expand the framework’s applicability.

\subsection{Implications of our Framework}\label{sec:III.G}

In spite of the aforementioned limitations, our WoT4EDP framework has significant implications for teaching and learning in STEM education, as well as educational research. By explicitly identifying and characterizing different forms of thinking—design, science, mathematical, metacognitive, and computational—this framework provides educators with a structured yet flexible approach to supporting and analyzing student thinking during engineering design-based tasks. For instance, instructors can use the framework to design activities that intentionally foster these thinking skills or assess student work with a focus on how well students integrate design and science thinking. The framework also offers guidance (see \textit{Enhancement approaches} in Table~\ref{tab:wot_concepts}) for scaffolding learning~\cite{dasgupta2019improvable, kolod_punt_2003}, enabling educators to better identify gaps in student understanding and adapt instruction accordingly.

Our studies leading to the development of the WoT4EDP framework primarily utilized data from group discussions and end-of-task written reports (outlined in Table~\ref{tab:data_earlier_studies}). The primary unit of analysis was the group, rather than individual students, with each group typically composed of three students (occasionally two), engaging in structured activities that yielded diverse insights. While written reports provided structured reflections, group discussions—though sometimes chaotic—revealed the dynamic, evolving nature of student thinking. Each group exhibited unique approaches, contributing to the richness of the data and demonstrating that no two groups processed or presented their ideas identically. Despite challenges in accurately attributing statements to specific speakers within the transcripts, this group-focused analysis has proven beneficial. The flexibility of the WoT4EDP framework allows it to capture ways of thinking across different time scales, from brief, momentary shifts to sustained engagement~\cite{sayre2015brief}, depending on task complexity. Incorporating multi-week written reports further highlights the framework’s adaptability to diverse data sources, showcasing its broad applicability in analyzing student engagement.

To demonstrate the utility of the WoT4EDP framework, consider a physics lab where students are tasked with designing a catapult~\cite{ravi_prper_2024, ravi_perc_2023}. Educators can apply the framework to assess not only how students conceptualize physics principles but also how they apply mathematical reasoning, engage in iterative design, reflect on their problem-solving strategies, and conduct relevant computational tasks using Python. 

At a practical level, instructors can use the WoT4EDP framework to develop rubrics~\cite{TEA2024, docktor2016assessing, huang2020developing, reynders2020rubrics} that assess student learning and performance in engineering design-based physics projects. The framework’s dimensions—such as design thinking, science thinking, mathematics application, and computational thinking (CT)—can be translated into measurable criteria for evaluating both the process and outcomes of student work. For instance, the rubric can assess creativity and feasibility in design thinking, depth of understanding and application of physics principles in science thinking, and the accuracy and relevance of mathematical and computational solutions to the problem. Computational thinking can be evaluated through criteria such as the logical structuring of algorithms, efficiency of code, or the integration of simulation tools to enhance design processes. Additionally, the framework's emphasis on integration allows instructors to assess how effectively students combine design, science, math, and computational tools to address real-life problems. By incorporating metacognitive reflection, instructors can encourage students to articulate their problem-solving strategies, computational approaches, and learning progress. This structured approach ensures consistency in assessment while offering students clear guidance on expectations.

Looking ahead, future research could explore how this framework applies across different disciplines and educational contexts, refining it to capture more subtle or discipline-specific nuances. One promising avenue for investigation is the potential integration of concepts from \textit{``Epistemological Framing''}—how students interpret and situate their learning experiences based on past experiences and knowledge~\cite{hutchison2010attending, irving2013transitions}. Additionally, exploring how factors such as group dynamics, instructional methods, and varying project types influence the emergence of these thinking patterns will help expand and refine the WoT4EDP framework.

While our previous studies have instilled confidence in the development of this framework, we acknowledge the importance of adhering to sound validation practices. We value Inglis' advice:
\begin{quote}
\textit{``It is important that developers of quality frameworks not fall into the trap of substituting intuition and guesswork for evidence-based validation processes.''~\cite[p.361]{inglis2008approaches}}
\end{quote}

Thus, validating the WoT4EDP framework will necessitate further empirical research. This may include case studies across diverse educational settings to assess its adaptability and robustness. Longitudinal studies tracking student progress over multiple projects could provide insights into whether the framework reliably captures growth in STEM thinking. Finally, examining other similar frameworks could offer valuable perspectives that enhance our understanding of effective validation practices, bringing us to the next section of this paper.


\renewcommand{\arraystretch}{1}
\begin{table*}[!htbp]
\begin{center}
\captionsetup{justification=raggedright} 
\caption{This table presents the \textit{Ways of Thinking} frameworks considered in this study. WoT4EDP refers to the proposed Ways of Thinking for Engineering Design-based Physics framework. FAWTEER, or the \textit{Framework for Applying Ways of Thinking in Engineering Education Research}~\cite{dalal2021developing}, is attributed to Dalal {\em et al.}. The \textit{Math and STEM-based Problem Solving} framework is referred to as English's Framework~\cite{english2023ways}.}

\begin{ruledtabular}
\label{fig:wot_comparison}
\begin{tabular}{p{0.25\linewidth} p{0.25\linewidth} p{0.45\linewidth}}
\textbf{WoT4EDP} & \textbf{Dalal {\em et al.}'s Framework}  & \textbf{English's Framework}\\
\hline
Design-Based Thinking   & Futures Thinking      & Critical Thinking (Includes Critical mathematical Modeling and Philosophical Inquiry)\\
Science-Based Thinking   & Strategic Thinking      & Systems thinking\\
Mathematics-Based Thinking   & Values Thinking      & Design-Based Thinking\\
Metacognitive Reflection   & Systems Thinking      & \\
Computational Thinking   &       & \\

\end{tabular}
\end{ruledtabular}
\end{center}
\vspace{-0.5cm}
\end{table*}

\section{W\texorpdfstring{\lowercase{o}}{o}T Frameworks - A Comparative Study}\label{sec:IV}

To address our second research goal \hyperref[RG2]{RG2}, this section examines two recently proposed Ways of Thinking frameworks that offer meaningful connections to our context. Section~\ref{sec:IV.A} provides a detailed exploration of Dalal {\em et al.}'s FAWTEER~\cite{dalal2021developing} framework, while Section~\ref{sec:IV.B} presents a comprehensive analysis of English's framework~\cite{english2023ways}. In each section, we not only discuss the context and elements of these frameworks, but also identify valuable insights and approaches that can inform and strengthen the development of our own WoT4EDP framework.

\subsection{Dalal {\em et al.}'s Framework} \label{sec:IV.A}

\renewcommand{\arraystretch}{1}
\begin{table*}[htbp] 
\centering 
\begin{center}
\captionsetup{justification=raggedright} 
\captionsetup{width=\textwidth}
\caption{The FAWTEER framework and its potential connections to the proposed WoT4EDP framework. Abbreviations: DBT—Design-Based Thinking; SBT—Science-Based Thinking; MBT—Mathematics-Based Thinking; MER—Metacognitive Reflection; CT—Computational Thinking.}

\label{tab:fawteer_edp}
\resizebox{\textwidth}{!}{%
\begin{ruledtabular}
\begin{tabular}{p{0.08\linewidth} p{0.16\linewidth} p{0.16\linewidth} p{0.16\linewidth} p{0.16\linewidth} p{0.16\linewidth}}
\textbf{FAWTEER} & \textbf{\hspace{1cm}DBT} & \textbf{\hspace{1cm}SBT} & \textbf{\hspace{1cm}MBT} & \textbf{\hspace{1cm}MER} & \textbf{\hspace{1cm}CT} \\
\hline
Systems Thinking   & How various components of the contraption or setup interact or influence each other. Identify constraints and stakeholders. Economic aspects. Invoking multiple disciplines.  & Identify systems and surroundings. Choose subsystems as appropriate. Apply physical laws or principles to various subsystems. & Apply mathematical models and concepts to various subsystems and integrate them. Use data-driven approaches. & Reflect on the challenges in integrating the multiple subsystems and adjust strategies accordingly. Raise new questions in the process. & Develop codes for simulating and automating processes and integrate them. Python code modules can break a complex problem into smaller manageable components. \\
Futures Thinking   & Envision long-term impacts of design decisions. Solutions may consider fostering long terms societal impacts. Engage in iterations based based on past learning. & Use of newer physics concepts to further technological innovation. & Use of mathematics and statistics to predict future trends, optimize future scenarios, and prepare for evolving system behaviours over time. & Brainstorm future scenarios, technologies, challenges, and innovations. & Simulate future scenarios by use computational models incorporating Machine Learning and Artificial Intelligence tools.\\
Values Thinking    & Consider ethical and user-centered designs. Ensuring solutions align with societal, environmental, and cultural values. Collaborate actively with stakeholders. & Consider ethical and responsible use of physics principles and ideas towards creating safe and sustainable solutions. & Display transparency in data collection, analysis, and decision making processes. Ensure the mathematical models factor in fairness and ethics. & Reflect on personal and societal values and be respectful to culture when making design decisions. & Ensure the codes or algorithms are not biased. Vouch for fair and responsible use of technological tools.  \\
Strategic Thinking & Consider criteria and constraints while planning a solution. Make optimal use of resources. Pay attention to economic aspects. Engage in collaborations.  & Make appropriate assumptions and approximations. Consider how various influencing factors may evolve over time. Divide the complex systems into subsystems and apply physics principles strategically. & Use mathematical models to predict future trends and guide decisions accordingly. & Conduct a feasibility study, discuss the effectiveness of the design or solution, and make adjustments as may be needed. & Use codes efficiently. Consider use of libraries, modules, functions etc. to develop a organized set of coding blocks. \\
\end{tabular}
\end{ruledtabular}
}
\end{center}
\vspace{-0.5cm}
\end{table*}

\subsubsection{Context}\label{sec:IV.A(i)}

The FAWTEER framework is specifically designed to address challenges in Engineering Education Research (EER). We see notable alignment between this framework and the context of engineering design-based (ED) physics laboratory projects, with broader relevance to Physics Education Research (PER). In our view, the framework offers valuable insights for both student learning and educational research, recognizing both students and educational researchers as critical stakeholders in the learning process. This perspective is further reinforced by Dalal {\em et al.}'s stated aim to \textit{``enact systemic changes''}~\cite[p.108]{dalal2021developing}, addressing how engineering education researchers think, make decisions, and engage in research to drive transformation within the system~\cite[p.110]{dalal2021developing}.

Among their motivations is \textit{The Research Agenda for the New Discipline of Engineering Education}~\cite{EER_2006} 2006 report, which identifies five research areas for the emerging discipline of Engineering Education, consisting of interrelated strands that can be explored independently or in conjunction with other fields. (i) \textit{Engineering Epistemologies} focuses on defining what constitutes engineering thinking and knowledge within current and future social contexts. (ii) \textit{Engineering Learning Mechanisms} investigates how engineering learners develop their knowledge and competencies within specific contexts. (iii) \textit{Engineering Learning Systems} examines how diverse human talents contribute solutions to social and global challenges relevant to the engineering profession. (iv) \textit{Engineering Diversity and Inclusiveness} explores the importance of equity and inclusion in engineering education. Finally, (v) \textit{Engineering Assessment} emphasizes the research and development of assessment methods, instruments, and metrics to enhance engineering education practice and learning. 

When viewed in conjunction with Bao and Koenig's~\cite{bao2019physics} \textit{Physics education research for 21st century learning}, notable connections are evident between the goals of Physics Education Research (PER) and Engineering Education Research (EER). Both fields emphasize the importance of fostering discipline-specific deep learning and enhancing scientific reasoning to facilitate knowledge transfer across STEM disciplines. Additionally, they highlight the necessity of promoting equity and inclusiveness, recognizing the value of diverse perspectives in enriching the educational experience. Furthermore, both EER and PER advocate for the research, development, assessment, and dissemination of effective educational practices, underscoring a shared commitment to improving teaching methodologies and student engagement in their respective fields~\cite{henderson2009impact, loui201911_EER, kautz2007physics, EER_2006, aapt2022strategic,mcdermott2001oersted, aseeVisionMission}.

\subsubsection{Framework Elements}\label{sec:IV.A(ii)}

FAWTEER emphasizes four specific \textit{Ways of Thinking}: \textit{Futures, Values, Systems}, and \textit{Strategic thinking}. It conceptualizes \textit{Way of Thinking} as a framework that can be practically applied to drive innovation and enhance researchers' capacity to effect systemic change. Notably, Dalal {\em et al.} acknowledge that other ways of thinking, such as entrepreneurial, computational, or design thinking, may also be relevant to their context. We resonate with their goal: to inspire further exploration of different ways of thinking in diverse contexts.

In the following discussion, we delineate each component of FAWTEER as articulated by Dalal {\em et al.} and illustrate each aspect with straightforward examples from our context. While we recognize that this may oversimplify the complexity of these concepts, our aim is to demonstrate how these elements can be effectively operationalized in simple everyday classroom tasks.

(1) \textit{Futures Thinking} involves exploring multiple possibilities for the future, focusing on addressing today's challenges to influence tomorrow. It emphasizes learning from past decisions, understanding the present, and anticipating consequences~\cite[p.110]{dalal2021developing}. In education, particularly in engineering design-based physics, futures thinking is crucial for fostering adaptability and innovation, preparing students for complex, evolving challenges. It aids in decision-making by envisioning multiple scenarios and preparing for potential outcomes. For instance, when designing a roller coaster, students can evaluate both short- and long-term impacts of their design choices, engage in iterative processes informed by past experiences, and emphasize the importance of continuous improvement while planning for potential disruptions.

(2) \textit{Systems Thinking} is a holistic approach that views a system as a cohesive entity composed of interconnected elements and subsystems. It emphasizes understanding interdependencies, structures, and functions while recognizing complexity, delays, and uncertainties~\cite[p.111]{dalal2021developing}. For instance, when students draw a free-body diagram—identifying all the forces acting on an object—they engage in systems thinking. This task necessitates decisions about defining the system and its surroundings. Similarly, when applying the principle of momentum conservation, students learn that while the total momentum of a system is conserved during a collision (assuming no external forces), the individual momentum of each body is not conserved. Although these examples may appear simplistic, they illustrate that even basic tasks can effectively convey the intricate concept of systems thinking to students, fostering deeper comprehension of complex interactions in physics. Our view is that educators must leverage the fact that there are \textit{``many different views regarding the definition of Systems Thinking, and as yet there does not seem to be a precise, widely-accepted definition''}~\cite{monat2015systems} to their advantage and further their educational goals. 

(3) \textit{Values Thinking} encompasses fundamental beliefs about right and wrong, and emphasizes ethical and normative considerations. It is about recognizing how decisions are made rather than determining their correctness, highlighting the importance of ethics, equity, and social justice within diverse cultural contexts. Values thinking may be viewed as self-reflection or introspection on one’s activities and assumptions, recognizing that perspectives may vary widely~\cite[p.110]{dalal2021developing}. If students design an experiment to address nuclear waste, they can engage in a values thinking exercise by reflecting on the ethical implications of their proposed solutions. They may be encouraged to consider questions such as, \textit{``Who may be affected by our solution?''}. This process underscores the importance of identifying stakeholders, promoting a deeper understanding of the social and ethical context surrounding their scientific work.

(4) \textit{Strategic Thinking} involves formulating a plan of action to achieve specific goals and is often regarded as a key leadership skill. It encompasses envisioning objectives, collaboratively developing plans, and effectively allocating resources to drive innovation. As a creative and ongoing process, strategic thinking translates the abilities of systems and futures thinking into actionable steps for change, enabling individuals to implement specific actions and achieve desired outcomes~\cite[p.111]{dalal2021developing}. In a design project to create a bulletproof vest, students apply strategic thinking by defining their goals of maximizing protection against bullets while ensuring comfort and mobility. They may start by researching the science behind bulletproof materials. Next, they may use mathematics to calculate optimal coverage areas and the weight distribution of materials to maintain comfort.  Python code may be employed to simulate the vest's performance under various impact scenarios, allowing them to visualize how different materials respond to bullet forces. By integrating design, science, math, and programming, they cultivate a comprehensive understanding of how strategic thinking can lead to innovative and effective solutions.

We considered how the elements of FAWTEER may align with the components of our framework, as detailed in Table~\ref{tab:fawteer_edp}. We acknowledge that the connections we have made may not be perfect, but we believe there is merit in this exploration. Each framework, while it receives our attention, holds its own unique significance.
We are reminded of James' philosophical perspective, which emphasizes the fluid and subjective nature of reality based on our experiences and interests, elegantly quoted by Schultz in his evocative essay  \textit{On Multiple Realities}: 
\begin{quote}
\textit{``Each world whilst it is attended to is real after its own fashion; only the reality lapses with the attention.''}~\cite[p.207]{schutz1962multiple}.
\end{quote}
Thus, our engagement with these frameworks contributes meaningfully to our understanding, even if the connections are not flawless.

\subsection{English's Framework}\label{sec:IV.B}

\renewcommand{\arraystretch}{1}
\begin{table*}[htbp] 
\centering 
\begin{center}
\captionsetup{justification=raggedright} 
\captionsetup{width=\textwidth}
\caption{English's SWoT framework for \textit{Mathematics and STEM-based Problem Solving} and its potential connections to the proposed WoT4EDP framework. These connections are flexible, encouraging educators to interpret and adapt these concepts to suit their specific contexts and goals. Abbreviations: DBT—Design-Based Thinking; SBT—Science-Based Thinking; MBT—Mathematics-Based Thinking; MER—Metacognitive Reflection; CT—Computational Thinking.}

\label{tab:english_edp}
\resizebox{\textwidth}{!}{%
\begin{ruledtabular}
\begin{tabular}{p{0.11\linewidth} p{0.16\linewidth} p{0.16\linewidth} p{0.16\linewidth} p{0.16\linewidth} p{0.16\linewidth}}
\textbf{English's Framework} & \textbf{\hspace{1cm}DBT} & \textbf{\hspace{1cm}SBT} & \textbf{\hspace{1cm}MBT} & \textbf{\hspace{1cm}MER} & \textbf{\hspace{1cm}CT} \\
\hline

Critical Thinking  & Students may articulate claims about their design decisions, supporting them with evidence derived from real-world data, physical properties, or theoretical principles. Their explanations are grounded in scientific (physics, in our context) concepts and principles. & Students construct science-based claims related to their designs, ensuring evidence and reasoning align with core scientific (physics) concepts and principles. & Students make claims (say, about optimal design performance), substantiating them with mathematical calculations rooted in scientific (physics) principles or logical reasoning. & Students engage in reflective discussions to evaluate how they developed their claims, the appropriateness of their evidence, and the alignment of their reasoning with their design goals, and scientific concepts and principles. & Students employ computational tools (say, by using Python code) to integrate physics principles and mathematics, automate calculations, and use visualizations (e.g., tables or graphs) to substantiate claims and explain their reasoning effectively. \\
Critical Mathematical Modelling  & Students evaluate design feasibility by developing mathematical models, questioning their assumptions, and considering the ethical impacts of their solutions on users and the environment. & Students explore scientific principles through modeling, analyzing how well their representations align with reality, and reflecting on the ethical implications of their findings.  & Students analyze relationships between variables using mathematical models, critically assessing their calculations and questioning the assumptions that influence real-world applications. & Students reflect on their modeling processes, evaluating their mathematical decisions and considering how these choices impact their design solutions and ethical considerations.  &  Students create simulations based on mathematical models, questioning the accuracy of their algorithms and considering the ethical responsibilities tied to their computational designs.  \\
Philosophical Inquiry & Students consider the ethical implications of their design choices, questioning how their solutions impact various stakeholders and society while prioritizing sustainability and inclusivity. & Students explore the moral responsibilities of scientists, reflecting on how their understanding of physical principles can lead to innovations that benefit or harm society. & Students evaluate the societal implications of their mathematical models, questioning how their calculations affect real-world applications and the potential consequences of inaccuracies.  & Students reflect on their thought processes, critically assessing their decision-making strategies and considering the ethical dimensions of their design and scientific choices. & Students analyze the ethical use of algorithms in their designs, reflecting on how computational models influence outcomes and the potential biases they may introduce. \\
\\

        \multicolumn{6}{p{0.9\textwidth}}{\textbf{Note:} \newline{Systems Thinking and Design-Based Thinking are not discussed here to avoid redundancy. See Tables~\ref{tab:wot_concepts} and ~\ref{tab:fawteer_edp}.}} \\
        
\end{tabular}
\end{ruledtabular}
}
\end{center}
\vspace{-0.5cm}
\end{table*}

\subsubsection{Context}\label{sec:IV.B(i)}

English's \textit{Ways of Thinking in STEM-based Problem Solving} framework appears to stem from an interest in understanding the cognitive processes that facilitate learning, problem-solving, decision-making, and interdisciplinary concept development. This framework directly connects to our own, especially when compared to the one proposed by Dalal {\em et al.} The emphasis on Mathematics within English's framework further strengthens this relationship. While a significant portion of English's research focuses on K-6 learners~\cite{english_profile}, her framework demonstrates broader applicability across various educational levels. 

\subsubsection{Framework Elements - Our Interpretation}\label{sec:IV.B(ii)}

The framework by English~\cite{english2023ways} comprises three main components: (i) \textit{Critical Thinking} (including critical Mathematical Modelling and Philosophical Inquiry), (ii) \textit{Systems Thinking}, and (iii) \textit{Design-based Thinking}. Interestingly, her inclusion of Systems Thinking creates a direct link to FAWTEER.

English argues that a focus on thinking skills can \textit{``ultimately lead to the development''}~\cite[p.1220]{english2023ways} of what she terms \textit{``Learning Innovation''}~\cite{mckenna2014adaptive}. However, we found some ambiguity in her treatment of this concept. In Table 1 and Section 3.4 of her article, \textit{Learning Innovation} is framed as a way of thinking~\cite[p.1222]{english2023ways}, whereas in Figure 1 (p. 1220), it appears positioned as a pedagogical goal~\cite[p.1220]{english2023ways}. While this inconsistency is somewhat perplexing, it does not present a significant issue for our analysis. We revisit the concept of \textit{Learning Innovation} in Section~\ref{sec:IV.B(iv)}.

To facilitate connections with our framework and for ease of discussions, we will treat \textit{Critical Thinking, Critical Mathematical Modelling, Philosophical Inquiry, Systems Thinking}, and \textit{Design-based Thinking} as distinct elements, while viewing \textit{Learning Innovation} as a pedagogical goal, despite her not making this explicit.  This approach will not hinder our analysis as our goal is to understand and draw insights from her framework.

\subsubsection{Framework Elements - Discussion}\label{sec:IV.B(iii)}

In the following discussion, we will outline each element of English's framework and provide simple examples from our context. Our aim is to emphasize the strong connections between our frameworks and demonstrate how educators can adapt these frameworks to meet their specific needs and objectives.

(1) \textit{Critical Thinking:} Several undergraduate STEM reform documents emphasize critical thinking as essential for developing and refining ideas, whether through explanations, designs, or investigations. Engaging in argumentation and critique foster this thinking and deepen understanding~\cite[p.125]{nrc_dber_2012}, ~\cite[p.xv]{NGSS2013}, \cite[p.46]{NRC2012}. Given this emphasis in reform documents, critical thinking rightly finds a place in English's framework. 

According to English, within STEM-based problem solving, critical thinking extends inquiry skills by involving the evaluation and judgment of problem situations. It encompasses assessing statements, claims, and propositions, analyzing arguments, drawing inferences, and reflecting on both solution approaches and the conclusions reached~\cite[p.1221]{english2023ways}. 

While research indicates that argumentation can enhance conceptual learning~\cite[p.1304, p.1305]{aydeniz2012argumentation},~\cite{pols2023integrating}, Osborne expresses concern about its notable absence in science education. He highlights the need for a deeper understanding of how argumentation facilitates learning and the characteristics of learning environments that foster effective student arguments~\cite{osborne2010arguing}. In our exploration of SWoT frameworks within scientific argumentation and evidence-based reasoning, we found Slavit {\em et al.}'s application of Toulmin's \textit{Claim-Evidence-Reasoning} (CER) framework~\cite{mcneill2009synergy, toulmin2003uses, pols2023integrating} in the context of integrated STEM to be particularly insightful~\cite[p.135]{slavit2022analytic},~\cite[p.464]{osborne2010arguing}. An examination of Slavit {\em et al.}'s analytic coding framework provides valuable insights with potential applications for future qualitative analyses, particularly in applying CER within integrated STEM contexts.

We found it beneficial to contemplate on the integration of critical thinking~\cite{wilson2020systematic} within our framework and context. For instance, in designing a roller coaster, students can effectively apply the \textit{Claim-Evidence-Reasoning} (CER) framework. They might start with the claim, \textit{``Our roller coaster will be safe and fun''}.  For evidence, they could use calculations using energy and momentum principles to show how their design meets safety regulations regarding speed, height, and weight. In the reasoning section, they would explain how these calculations ensure an enjoyable ride without compromising safety. 

(2) \textit{Critical Mathematical Modelling:} We were intrigued by English's preference for the term \textit{critical mathematical modelling} over simply \textit{mathematical modelling}. She defines modeling as the development of conceptual innovations that address real-world needs, emphasizing that effective modeling must include contextual understanding and critical analysis. Moreover, she argues that critiquing the outcomes of a model and reflecting on the lessons learned is crucial when addressing real-world problems~\cite[p.1222]{english2023ways}. 

English's approach is echoed by Stillman {\em et al.}, who stress the importance of learning environments that enhance students' social-critical efficacy. They promote a socio-critical mathematical modeling cycle, allowing students to address problems relevant to their interests and communities while producing solutions that are socially significant~\cite[p.18]{math_model_stillman_rosa}. This perspective aligns with current educational practices that emphasize real-world applications and social relevance in mathematics.

In our context, as an example, a catapult design project enables students to collaborate in launching a projectile to a specific range while applying principles of projectile motion. This activity incorporates mathematical modeling, allowing students to make assumptions about factors such as launch angles, launch speed, and drag force. They engage in critical discussions regarding material choices and design constraints, fostering argumentation as they justify their decisions. Before creating an actual model, students can run simulations in Python to measure projectile range, which facilitates reflection and iteration as needed. Furthermore, showcasing their models at a community event not only makes the project fun and educational but also extends the reach of science to the broader community.

(3) \textit{Philosophical Inquiry:} In our framework, we designate philosophical inquiry as a pedagogical goal, while English advocates for it as a way of thinking. As previously mentioned, having diverse perspectives is not problematic but indeed enriching. English argues that philosophical inquiry encompasses various thinking skills, including identifying hidden assumptions, exploring alternative actions, and reflecting on conclusions and claims. We align with her perspective that such inquiry can foster conceptual depth~\cite[p.1223]{english2023ways}.

Our laboratory work emphasizes group collaboration, and a significant portion of our research data is derived from group discussions. We appreciate English's assertion that philosophical inquiry facilitates group problem-solving, peer-sharing of created models, and constructive feedback. In a recent computational physics laboratory activity, our students analyzed actual \textit{big data} from a High Energy Physics research laboratory~\cite{glover2024cms}. Inspired by English's framework, in our next iteration, we see scope to guide our students to pose critical questions, develop models, and engage in meaningful debates about their conclusions on the same problem. 

In this context, Gardner implores us to get comfortable with the truth, emphasizing that true inquiry transcends mere conversation. She cautions educators to differentiate between facilitating dialogue and facilitating genuine inquiry, noting that quite often \textit{``just dialogue continues to pass for philosophical inquiry''}~\cite[p.72]{gardner2015inquiry}. This distinction has prompted us to reflect on our practices critically.

While our framework already incorporates metacognitive reflection as a way of thinking, we recognize that engaging in reflective thinking and \textit{``free-flowing''}~\cite{wu2009improving, ravi_prper_2024} conversations can further enhance philosophical inquiry. This integration could deepen students' understanding and foster a richer educational experience.

(4) \textit{Systems thinking:} We observe notable parallels between Dalal {\em et al.}'s and English's discussions on systems thinking. To minimize repetition, we highlight key ideas. English views systems thinking to encompass consideration of system boundaries, system  components, the interactions among the components, and the emergent properties and behavior of the system~\cite[p.1222]{english2023ways}. Consistent with our view, she also acknowledges that systems thinking overlaps with other ways of thinking. Particularly notable is her concern that systems
thinking is almost absent from mathematics education, despite its importance in STEM contexts. This prompted us to explore existing literature, where we found at least one study utilizing systems thinking as a lens to examine students' understanding of word problems in school mathematics~\cite{salado2019systems}.

In our context of an undergraduate physics classroom, while we do not always explicitly emphasize systems thinking, we indirectly cover its aspects. For instance, in mechanics, when teaching conservation laws, we can prompt students to define their system and specify the conditions under which these laws apply. Similarly, in multi-loop electrical networks, we can encourage students to consider which voltage or current source influences the current in a particular branch. We can ask them how many equations are necessary to solve for all currents in the network, given the voltages and resistances, and discuss how changes to one circuit component can affect the entire system.

Furthermore, when students write multiple cells of Python code, they can explore how the program's output changes when running specific cells. This approach fosters an understanding of the interdependence within the code structure, illustrating systems thinking across various domains.

(5) \textit{Design-based Thinking:} To minimize repetition, we summarize only a few key ideas. English defines design-based thinking as encompassing iterative problem scoping, idea generation, designing, solving, testing, reflecting, and finally communicating~\cite[p.1222]{english2023ways}. In the context of our engineering design-based physics projects, where the problem itself serves as a natural setting not only for applying known physics concepts but also for facilitating new learning, we find ourselves resonating with English's perspective that design-based thinking fosters \textit{``learning while designing''}, which she describes as \textit{``generative learning''}. Equally noteworthy is her assertion that design-based thinking, alongside other ways of thinking, can significantly contribute to \textit{Learning Innovation}~\cite[p.1225, 1226]{english2023ways}.

Finally, in a similar approach to our analysis of Dalal {\em et al.}'s framework, we closely examined the alignment between the elements of English's framework and the components of our proposed framework. This comparison allowed us to identify potential intersections, which are summarized in Table~\ref{tab:english_edp}. These insights contribute to our understanding of how various frameworks can inform and complement our model in the context of engineering design in STEM education.

\subsubsection{A Note on Learning Innovation}\label{sec:IV.B(iv)}

According to English, \textit{`Learning Innovation'} involves leveraging foundational or core disciplinary content knowledge to foster a more sophisticated understanding within the discipline. Drawing on insights from Berkun's \textit{The Myth of Innovation}~\cite{berkun2010myths} and acknowledging that there may not be a singular definition or interpretation of the term `innovation'~\cite{stenberg2017does}, we advocate for a mindful approach to the term, considering the complexities it entails.

Educators should feel empowered to interpret `innovation' in ways that best align with their local contexts and objectives. In our view, `innovation' is not merely about dramatic breakthroughs; rather, it often involves steady, incremental, and iterative improvements. This perspective aligns with Berkun's assertion in his thought-provoking essay \textit{Why Innovation is Overrated}~\cite{berkun_innovation_overrated} and that genuine innovation arises from action and problem-solving rather than the pursuit of buzzwords.

\subsection{Limitations of the Comparative Study}\label{sec:IV.C}
A few limitations merit discussion. First, in comparing WoT4EDP with other frameworks, we may have placed greater emphasis on aspects that are particularly relevant to our specific context, potentially overlooking important considerations for other disciplines. Second, our focus primarily on undergraduate introductory physics students may have introduced an unintended bias in our analysis. Finally, while we strongly advocate for flexibility in applying any framework, it is possible that some educators may prefer a more structured approach.

\subsection{Implications of the Comparative Study}\label{sec:IV.D}

This study has implications for various stakeholders, including teachers, students, and educational researchers. Notwithstanding the fact that the primary focus of this study is SWoT frameworks, our findings also contribute to the broader study of WoT frameworks.

The WoT framework can aid teachers in crafting detailed rubrics for a variety of projects. By exploring various WoT frameworks, teachers can enhance their rubrics, incorporating diverse perspectives to better evaluate and support student performance. Utilizing WoT frameworks provides valuable insights into student thinking, enabling teachers assess students’ problem-solving abilities.

Awareness of WoT frameworks can support students in reflecting on and refining their thinking skills. Rubrics based on these frameworks clarify evaluation criteria and expectations, guiding students to focus on crucial elements of their work and improve their performance.

Educational researchers may explore the process of developing and validating new WoT frameworks, contributing to the broader body of knowledge in educational settings. Engaging in a comparative analysis of different frameworks can provide insights into the strengths and limitations of different approaches. Investigating students' thinking guided by WoT frameworks can inform the creation of targeted pedagogical interventions and strategies to address specific aspects of students' thinking and problem-solving skills.

\section{Summary}

This study was driven by two main objectives: \hyperref[RG1]{RG1} focused primarily on our proposed WoT4EDP framework, while \hyperref[RG2]{RG2} aimed at comparing it with two other recently proposed frameworks.

In response to \hyperref[RG1]{RG1}, we introduced our WoT4EDP framework (see Figure~\ref{fig:WoT_framework}) as a part of an ongoing and evolving discussion on \textit{STEM Ways of Thinking} SWoT frameworks. In sections~\ref{sec:III.A} and~\ref{sec:III.B}, we detailed the context of the Engineering Design-based physics problems within an introductory physics laboratory. Section~\ref{sec:III.C} outlined our prior qualitative studies (see Table~\ref{tab:data_earlier_studies}), placing us in a strong position to contribute to the literature on SWoT frameworks. Notably, we introduced computational thinking as a crucial element missing in other frameworks. Section~\ref{sec:III.D} characterized the five core elements of our framework, supported by reform documents and relevant literature.  Table~\ref{tab:wot_concepts} provided a basic operational scheme for applying the framework, emphasizing flexibility for practitioners. We believe that designing classroom instruction with a deliberate focus on the SWoT elements can help achieve diverse pedagogical goals, presented in a flexible and non-exhaustive list inspired by our experience as educators and supported by reform documents (see section~\ref{sec:III.E}). We acknowledged the limitations of our comparative study in section~\ref{sec:III.F} and discussed the practical implications of our framework, highlighting simple yet practical benefits for students, teachers, and educators in section~\ref{sec:III.G}. 

To address \hyperref[RG2]{RG2}, we conducted an in-depth comparison between our framework and two others: one by Dalal {\em et al.} in the context of Engineering Education Research (EER), and the other by English, focusing on Mathematical and STEM-based problem solving. We outlined the context of each of the studies in sections~\ref{sec:IV.A(i)} and~\ref{sec:IV.B(i)}, and elaborated on each framework in sections~\ref{sec:IV.A(ii)} and~\ref{sec:IV.B(iii)} respectively. It was insightful and illuminating to draw connections between our framework and the other two, particularly in the comparative analyses shown in Tables~\ref{tab:fawteer_edp} and~\ref{tab:english_edp}. While acknowledging the limitations of our comparative analysis in section~\ref{sec:IV.C}, we also identified a few key implications of our comparative study in section~\ref{sec:IV.D}. 

Throughout our study, \textit{Utilization-Focused Evaluation} (see section~\ref{sec:II.F}) served as a valuable companion, helping us gain a deeper understanding of other frameworks while critically reflecting on our own. We have been both enlightened and humbled by this process of learning.

\section{Future Work}
In the second installment of our two-part series, we will demonstrate the application of the WoT4EDP framework as an analytical lens to examine new data from student group discussions and written reports in student-generated engineering design projects. We will quote verbatim at least two examples of physics-based ED problems generated by students. This study would explore how students engage with design, science, mathematics, and metacognitive reflection throughout their problem-solving processes. Additionally, we investigate students’ demonstrations of computational thinking—particularly in their Python coding—as they tackle real-world challenges. We would like to add a caveat that the data was collected in Spring 2023, well before the lead author encountered the term SWoT frameworks (see Section~\ref{sec:II.E}). 

A key challenge in qualitative research is the issue of overlapping codes, which we addressed extensively in our recent paper~\cite{ravi_prper_2024}. A common approach to qualitative analysis involves segmenting textual data, yet given the complexity of students’ reasoning and expression, multiple ways of thinking often co-occur within a single statement. While we previously illustrated such code co-occurrences with examples, qualitative analysis extends beyond mere segmentation. Over-segmentation, in particular, risks fragmenting students’ thought processes, raising critical questions about how best to delineate and interpret their ideas. We explored these issues in depth in our prior work and will expand on this discussion in the forthcoming study~\cite[p.7-10]{ravi_prper_2024}.

In the follow-up study, the complexity of the data increases as we analyze not only students’ written and verbal responses but also their Python code, images, graphs, and tables. Consequently, our analysis requires a more nuanced approach beyond textual segmentation. To address this, we employ a multi-pronged strategy integrating traditional \textit{coding}~\cite{saldana2009introduction}, \textit{thematic analysis}~\cite{braun2006using}, and \textit{thick description}~\cite{stahl2020expanding, younas2023proposing, brink1987reliability}. We will revisit these challenges in the follow-up paper, providing explicit details on our data parsing methods and coding sheets.

While we present one approach to analyzing this complex dataset, we acknowledge that different researchers may adopt varied methodologies depending on their research goals and interests. By applying the WoT4EDP framework, we aim to generate insights into students’ \textit{Integrated Thinking} and \textit{Learning Innovation}, as emphasized by English. This study not only extends the application of our framework but also contributes to the broader discourse in STEM education on effective teaching and learning practices. In the process, we hope to contribute to the discussion on the use of qualitative methods in PER. 

\section{Acknowledgments}
ChatGPT-4o and Perplexity were employed by the lead author to \textit{wordsmith} passages, and as a \textit{learning partner}. We gratefully acknowledge the insightful and encouraging comments of anonymous referees on our earlier publications and the current manuscript, which contributed to what we believe is a more coherent, detailed, and well-structured paper. We owe our gratitude to all the authors referenced, for they provided us with innumerable moments of much needed inspiration. A special thanks to Amir Bralin for designing the learning materials, with particular emphasis on scaffolding the Python-based activities and preparing the Jupyter Notebooks for the laboratory reports. We deeply appreciate the highly thoughtful insights Dr. Amogh Sirnoorkar offered towards the preparation of this manuscript. The lead author takes full responsibility for any error or omission. 

This work is supported in part by U.S. National Science Foundation grant DUE-2021389. Opinions expressed are of the authors and not of the Foundation.
\clearpage
\bibliography{references}
\clearpage
\appendix
\onecolumngrid

\appendix
\onecolumngrid

\end{document}